\documentclass{aa}
\usepackage{graphicx}
\usepackage{natbib}
\bibpunct{(}{)}{;}{a}{}{,}
\begin{document}
\title{The isolated interacting galaxy pair NGC 5426/27 (Arp 271)}

\author{I. Fuentes-Carrera\inst{1}
\and M. Rosado \inst{1}
\and P. Amram \inst{2}
\and D. Dultzin-Hacyan \inst{1}
\and I. Cruz-Gonz\'alez \inst{1}
\and H. Salo \inst{3}
\and E. Laurikainen \inst{3}
\and A. Bernal \inst{1}
\and P. Ambrocio-Cruz \inst{1}
\and E. Le Coarer\inst{4} }

\offprints{I. Fuentes-Carrera,  \email{iqui@astroscu.unam.mx}}

   \institute{Instituto de Astronom\' \i a,
Universidad Nacional Aut\'onoma de M\'exico (UNAM),
  Apdo. Postal 70-264, 04510,
M\'exico, D.F., M\'exico \\
\and
Laboratoire d'Astrophysique de Marseille, 2 Place Le Verrier,
Marseille Cedex 4, France \\
\and
Department of Physical Sciences, Division of Astronomy,
University of Oulu, FIN-90570, Oulu, Finland \\
\and
Observatoire de Grenoble, B.P.53X, F-38041, Grenoble Cedex 9, France \\
             }

   \date{Received . . . . . . . . . .  ; accepted . . . . . . . . . . }

   \abstract{
We present  $H\alpha$ observations of the isolated interacting galaxy pair NGC 5426/27  using the scanning  Fabry-Perot interferometer PUMA.  The velocity field, various kinematical parameters and rotation curve for each galaxy were derived. The FWHM  map and the residual velocities map were also computed to study the role of non-circular motions of the gas. Most of these motions can be associated with  the presence of spiral arms and structure such as  central bars. We found a small bar-like structure in NGC 5426, a distorted velocity field for NGC 5427 and a bridge-like feature between both galaxies which seems to be associated with NGC 5426.  Using the observed rotation curves,  a range of possible masses was computed for each galaxy.  These were compared with  the orbital mass of the pair derived from the relative motion of the participants. The rotation curve of each galaxy  was also used to fit different mass distribution models  considering the most common theoretical dark halo models. An analysis of the interaction process  is presented and a possible 3D scenario for this encounter is also suggested.

\keywords{galaxies: interactions --- galaxies: kinematics and dynamics --- galaxies: individual
 (NGC 5426, NGC 5427) --- galaxies: spiral} 
}
\authorrunning{Fuentes-Carrrera et al}
\titlerunning{Interacting galaxy pair Arp 271}
 \maketitle

\section{Introduction}

Galaxies suffer  modifications  from close
companions from  the  early  stages  of their  lives. Encounters
between galaxies are  common and important phenomena in the
evolution and transformation  of these systems. The way the
interacting process between two galaxies triggers, sustains or
inhibits the formation of structure in spiral galaxies has been
studied from different points of view, from broad statistical
studies to more detailed morphological, photometric and
kinematical analysis.  Several attempts have been made to
kinematically constraint the evolutionary scenario  of
interactions  and the formation of structure such as  bars, tidal
arms,  amongst others
\citep{salo93,saloa00,salob00, miw98, kauf99}. It has been found
that although morphology is relatively easy to reproduce with
numerical simulations \citep{toom72, barn88, barn96}, kinematical 
information imposes important
restrictions on  the models applied \citep{salo93, saloa00,
salob00}. 
In these works, N-body simulations have been compared with observations. 
The predicted shape of the rotation curve, deviations of the  gaseous spiral arms 
from the plane of the galactic disk,
peculiar high velocities in certain directions of the galaxies and
non-circular motions are constrained by 2D velocity fields produced by Fabry-Perot observations.

From the observational point of view, most of the work on the
kinematics and dynamics of interacting galaxy pairs has been done
using  long-slit spectroscopy along certain predetermined
positions thus restraining the kinematical information to a few
points on the galaxy. Nevertheless, for a perturbed system, it is important to
have  kinematical information of larger portions of the disk using
observational techniques such as extended-object Fabry-Perot
scanning  interferometry   since extended kinematical  information can help  
determine the effect the interacting
process has had on each of the members of the interacting system  and because axial 
symmetry is probably lost during the interaction  \citep{marc87, amr94, amr02}.

The study  of the dynamics of binary systems also allows one to estimate the total 
mass of a galaxy
as well as its  mass-to-luminosity ratio ($M/L$).  Through this
method, the integrated mass of a component can be  measured out to
10 times the luminous radius of an isolated galaxy. In an
elaborate statistical  study to determine $(M/L)$ in the V band for  binary
galaxies, \citet{schb87} found  that in general the   values for
Sc galaxies in interacting pairs range from 12 to 32, with a most
likely value of   $21 \pm 5$  in  solar units. Other works
suggest that for this type of  galaxy, the value of  $(M/L)$ in the B band  lies
between 10 and 16 \citep{cheng95, nord98, hon99}.
Most of the determinations of $ M/L$  for  binary galaxies are
based on statistical studies of very large samples; they thus
depend on the selection criteria of the latter as well as on the
statistical  method chosen to probe the $M/L$  of modeled pairs
versus observed pairs.  It is then of  special interest to focus
on individual pairs and attempt to determine their $M/L$ ratio through 
detailed observations and analysis  of their kinematics and
dynamics.
These observations should be complemented by  numerical simulations of the
encounters varying such parameters as $ M/L $  and the structure
of the dark matter halo while imposing  kinematical restrictions obtained
from interferometric 3D observations.

In this work, we present scanning Fabry-Perot  observations and the kinematical  
analysis of the
interacting galaxy pair NGC 5426/27 (Arp 271). Section 2 is an
overview of the observational parameters and the reduction process.
In section 3,  a brief  bibliographical review of recent works and
results  on this interacting galaxy pair is presented.
 In section 4,  we present the derived  velocity  field and the associated rotation curve of each
galaxy.
The kinematics inferred from the velocity fields and rotation
 curves are discussed in section 5,  giving special attention to the role
 of non-circular motions. In section 6 the mass of each galaxy is estimated
 through different methods as well as  its mass-to-light ratio.
Some suggestions on the structure of the dark matter halo are
also presented. In section 7, a possible 3D configuration for the
encounter is presented, several features which seem to have been
triggered or enhanced by the interaction process are also
discussed. Our conclusions are presented in section 8.

\section{Observations}

\subsection{Observations and Data Reductions}

Observations of NGC 5426/27 (Arp 271) were done on May 1997 at the f/7.5 Cassegrain focus of  the 2.1 m
 telescope at the Observatorio Astron\'omico Nacional in San Pedro M\'artir (M\'exico), using the
 scanning Fabry-Perot interferometer PUMA \citep{ros95}.

We used a $1024 \times 1024$ Tektronix CCD detector.
In order to enhance the signal, we applied a $2 \times 2 $ binning in both
 spatial dimensions so that the resulting image format was of $ 512 \times 512$  pixels  with  a
spatial sampling of  $1.18  \ \arcsec /  pixel$. In order to isolate the redshifted
 $ H \alpha \ (\lambda_{at \  rest} = 6562.73 \ \AA ) $  emission, we used an interference filter
centered at  $   6650 \ \AA $ with a  $ FWHM$ of   $ 47 \AA $.

PUMA is a focal reducer built at the Instituto de Astronom\'\i a-UNAM used 
to make direct imagery and Fabry-Perot  (FP) spectroscopy  of extended
emission sources (field of view $10 \arcmin $). The FP used is an
ET-50 (Queensgate Instruments) with a servostabilization system
having  a free spectral range of $19.95  \  \AA $ ($912 \ km \
s^{-1} $) at $H\alpha$. Its finesse ($ \sim 24$) leads to a
sampling spectral resolution
 of $0.41 \ \AA \ (19.0\  km \  s^{-1})$  which is achieved by scanning the
interferometer free spectral range through 48 different  channels.

To average the sky variations during the exposure, we got two data
cubes with an exposure time of 48 minutes each (60 s per channel).
 These data cubes were co-added leading to a total exposure time
of  96 minutes. For the calibration we used a He lamp whose
$6678.15 \ \AA $ line was close to the redshifted nebular
wavelength. Two calibration cubes were  obtained at the beginning
and at the end to check the metrology.

The instrumental and observational parameters are shown in Table 1.

\begin{table}
     \caption[]{Instrumental and Observational Parameters }
     $$
         \begin{array}{p{0.5\linewidth}l}
\hline
            \noalign{\smallskip}
           Parameter     &   Value    \\
            \noalign{\smallskip}
\hline
            \noalign{\smallskip}
           Telescope  &   2.1 \ m \  \ (OAN, \ SPM) \\
           Instrument &   PUMA  \\
           Detector &  Tektronic \ CCD  \\
            Detector size   &   (1024 \ \times \ 1024) \ px  \\ 
           Image scale (after 2x2 binning) &  1.18  \ \arcsec \ per \ px  \\           
           Scanning F-P interferometer &  ET-50 \ (Queensgate) \\
           F-P interference order  at $H\alpha$    &  330  \\ 
           Free spectral range at  $H\alpha$   &  19.95 \ \AA \  \ (912 \ km/s) \\
           Spectral sampling at  $H\alpha$  &   0.41 \ \AA  \ \ (19.0 \ km/s) \\
           Interference filter & 6650 \ \AA \  (FWHM=47 \ \AA)\\    
           Total exposure time  &  96 \ min   \\
          Exposure time (direct image) &  120 \ s  \\
          Calibration line  & 6678.15 \AA  \ (He \ lamp) \\      
            \noalign{\smallskip}
\hline
         \end{array}
     $$

   \end{table}

\begin{figure*}
\centering
\includegraphics{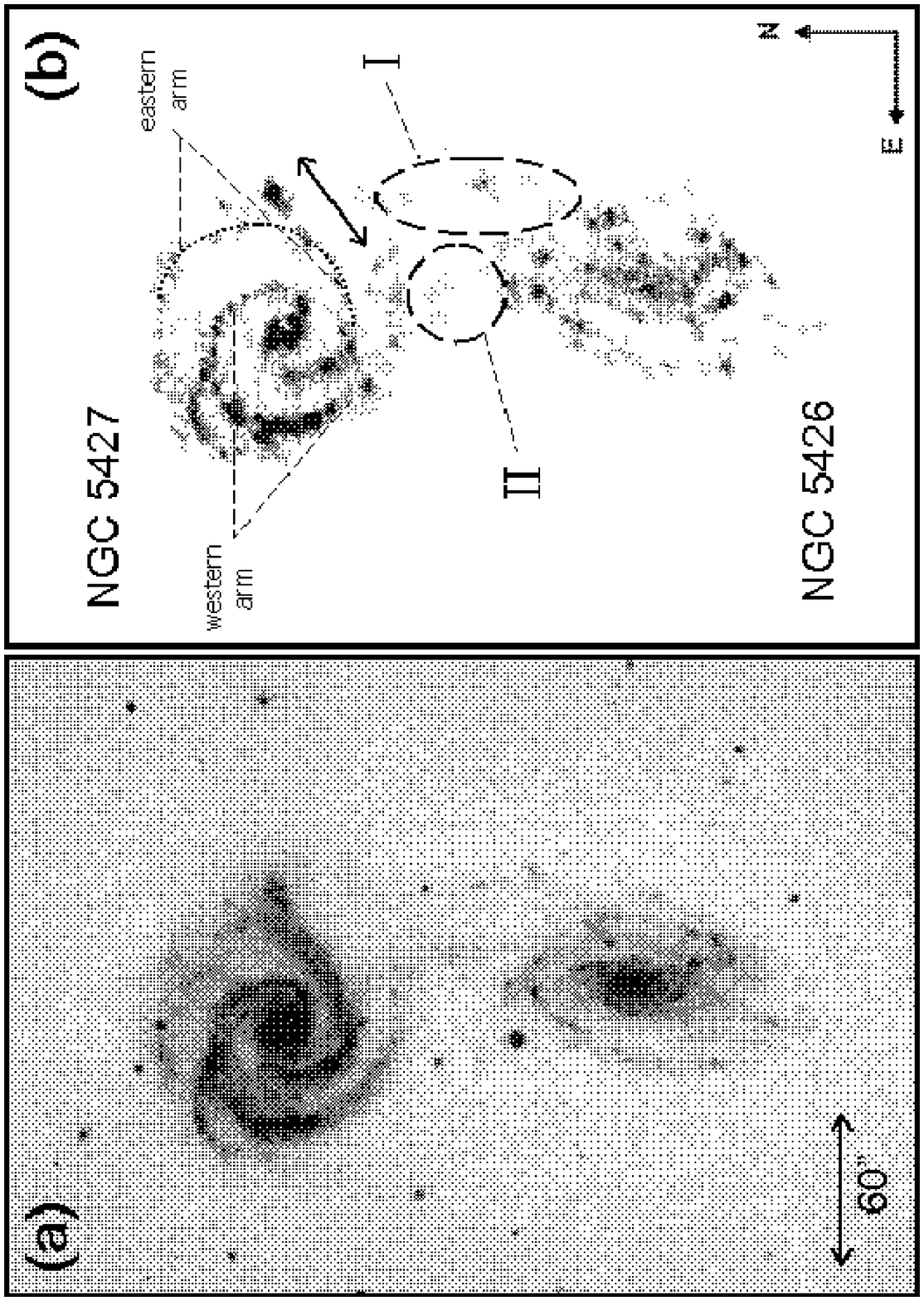}
\caption{a) Direct B image  of  NGC 5426/27 (Arp 271) taken from 
``The Carnegie Atlas of Galaxies. Volume II'' \citep{sand94}. 
b) Monochromatic $H{\alpha}$  (continuum substracted) image  of the pair obtained with the scanning Fabry-Perot interferometer PUMA data cubes (see text for reduction process). 
Arrow indicates straight-arm segment, dotted line traces what would be a ``classical'' spiral arm.
Regions I and II are associated to the bridge-like feature of gas between both galaxies  }
\end{figure*}

\subsection{Data Analysis}

The data reduction and analysis were done using mainly the
ADHOCw\footnote{{\it{http://www.oamp.fr/adhoc/adhocw.htm} developed by J. Boulesteix}}
software and the CIGALE software \citep{lec93}.
Standard corrections were done on each of the cubes:
removal of cosmic rays,  bias and dark subtraction, and
flat-fielding. Once the object  cubes were co-added, the night sky
continuum and OH sky lines ($6627.2 \ \AA$ and $6634.2 \ \AA$)
were subtracted. A spectral Gaussian smoothing ($\sigma = 57  \ km \
s^{-1}$) was also performed. Once the spectral smoothing was done, 
the calibration in wavelength was
fixed for each profile at each pixel  using the calibration cube.

Through the scanning process, we obtained for each pixel a flux
value at each of the 48 scanning steps. The intensity profile
found along the scanning process  contains information about the
monochromatic emission ($H\alpha$)  and the continuum emission of
the object. The continuum image computation was done considering
the mean of the 3 lowest intensities of the 48 channels cube. For
the monochromatic image, the $H \alpha $ line intensity was
obtained by integrating the monochromatic profile in each pixel.
The velocity maps were computed using the barycenter of the
$H\alpha$ profile peaks at each pixel. To get a sufficient
signal-to-noise ratio on the outer parts of each galaxy, we
performed three spatial Gaussian smoothings ($\sigma = 2.36, \
3.54, \ 4.72 \ \arcsec$) on the resulting calibrated cube. A
variable-resolution radial velocity map was built using high
spatial resolution (less spatially-smoothed pixels) for regions
with originally higher signal-to-noise ratio. No double peaks
appeared in the velocity profiles.

The profiles were deconvolved by the instrumental profile.
The $FWHM$  for each pixel was computed from
 the deconvolved profile at each pixel.
For each pixel, the threshold  of the profile  was defined by the value of the channel 
for which $60 \%$ of the channels had a larger intensity value.

\begin{flushleft}
\begin{table*}
\centering
     \caption[]{Parameters of NGC 5426 and NGC 5427 }
 \begin{tabular*}{1\linewidth} [  ] {l c c c c}
\hline
 \noalign{\smallskip}
              & \ \ \ \ \ \ \ \ \ \   &    NGC 5426  & \ \ \ \ \ \ \ \  \ \ \ \ \ \ \ \ \ \ \ &  NGC 5427  \\
\noalign{\smallskip}
\hline
 \noalign{\smallskip}
           Coordinates  (J2000)$^{\mathrm{a}}$   &  &     $\alpha = 14h \ 03m \ 24.8s  $    &   &  $\alpha = 14h \ 03m \ 26.0s  $   \\
                                        &    &      $\delta = - 06^\circ \ 04\arcmin \ 09\arcsec    $    &    & $\delta = -06 ^\circ \ 01\arcmin \ 51\arcsec   $  \\

          Morphological type$^{\mathrm{b}}$  &  &  SA(s)c pec &  &  SA(s)c pec Sey 2 \\

         Luminosity class$^{\mathrm{c}}$ &    &      II     &  &    I \\

         $m_{B}$$^{\mathrm{d}}$ (mag) & &  $12.86$  & &  $12.06$ \\

         $D_{25}   /   2$$^{\mathrm{d}}$ ($\arcmin $)& &  $1.475$ & & $1.435 $ \\

        Distance$^{\mathrm{e}}$ (Mpc)  & &  - - -     &  &26.7 \\

        $M_{B}$$^{\mathrm{d}}$ (mag) &  &$- 20.567$ & &$- 21.167$ \\

         Average surface brightness  within  $D_{25}   /   2$$^{\mathrm{d}}$ (in $mag/\arcsec^2$) & &$23.50 $ & &$22.86$  \\

     Systemic velocity \ \ $( km/s) $ &  &$2516 \pm 5$$^{\mathrm{c}}$ &  &$2703 \pm 20$$^{\mathrm{c}}$ \\

                                                                         &  &$2584 \pm 9$$^{\mathrm{f}}$ &  &$2730 \pm 9$$^{\mathrm{f}}$ \\

                                                                          &  &$2575 \pm 3$$^{\mathrm{g}}$ &  &$2722.5 \pm 1$$^{\mathrm{g}}$ \\

          $ V  _{rot \ max}$ \ \  $(km/s)$& & $209$$^{\mathrm{h}}$ &  &  $172$$^{\mathrm{h}}$  \\

                                                                                                                    & & $200$$^{\mathrm{i}}$ &  &   - - -   \\

              P.A. $( ^\circ )$ & &$180$$^{\mathrm{c}}$   &  &$68$$^{\mathrm{c}}$  \\

                                                & &$177.5 \pm 1 $$^{\mathrm{g}}$   &  &$53.2 \pm 3$ $^{\mathrm{g}}$ \\

           Inclination  $( ^\circ )$ &  &$59 \pm 2$$^{\mathrm{c}}$  & &$32 \pm 2$$^{\mathrm{c}}$ \\

                                                               & &            - - -                                &  &$24.49$$^{\mathrm{j}} $ \\

                                                              &  &$59 \pm 3$$^{\mathrm{g}}$  & &$34 \pm 2.5$$^{\mathrm{g}}$ \\

 Mass (in $ 10^{10} \ M_\odot$) &  & $3.39 \pm 0.13$$^{\mathrm{c,k}}$  & &$3.1 \pm 0.4$$^{\mathrm{c,l}}$ \\

                                                              &  &$11.2$$^{\mathrm{g,m}}$  & &$7.5$$^{\mathrm{g,m}}$ \\

\noalign{\smallskip}
\hline
\end{tabular*}
  
\begin{list}{}{}
\item[$^{\mathrm{a}}$] NED database
\item[$^{\mathrm{b}}$] de Vaucouleurs et al 1991 
\item[$^{\mathrm{c}}$] Blackman 1982
\item[$^{\mathrm{d}}$] LEDA database
\item[$^{\mathrm{e}}$] Through tertiary indicators (de Vaucouleurs 1979)
\item[$^{\mathrm{f}}$] Schweizer 1987a
\item[$^{\mathrm{g}}$] This work
\item[$^{\mathrm{h}}$] From $H\alpha$ rotation curve -this work 
\item[$^{\mathrm{i}}$] From 21 cm observations (Bottinelli et al 1948)
\item[$^{\mathrm{j}}$] Keel 1996
\item[$^{\mathrm{k}}$] Inside $7.8 \ kpc = 0.84 \ D_{25}/2$ using method of Nordsieck (1973) 
\item[$^{\mathrm{l}}$] Inside $9.4 \ kpc = 0.68 \ D_{25}/2$ using method of Nordsieck (1973) 
\item[$^{\mathrm{m}}$] Inside $D_{25}/2$ considering  a spheroidal distribution
\end{list}

   \end{table*}
\end{flushleft}

\section{NGC 5426 and NGC 5427}

Arp 271  is an interacting pair consisting of two spiral galaxies, 
NGC 5426  and NGC 5427,
of approximately the same angular size ($2.3  \arcmin$)
(Figure 1). This pair was first catalogued by \citet{vor59}. It
was later  included in Arp's {\it Atlas of Peculiar Galaxies}
 \citep{arp66} grouped with other objects he classified as double
 galaxies with connecting arms. This connection is clearly visible in the
ultra-high contrast J-band  image of the pair   presented  by
 \citet{bla82}.

 Several works have been done  on this pair of galaxies,  mostly as part
 of vast surveys of interacting galaxies. In an extensive analysis of morphological
correlation for over 16,000 nearest neighbor pairs of galaxies,
 \citet{yam89} found that NGC  5426 and NGC 5427 not only have the same morphological type
 but also very similar internal structure as well as comparable sizes.  These characteristics make Arp 271
fall into the category of {\it twin galaxies} proposed by these authors. According to them, this type of paired
galaxies suggest the possibility of fission processes  \citep{page75}  at the origin of the present
 morphological type of galaxies since it is easy to understand this  morphological resemblance  if the galaxies
considered formed under the same initial conditions given by the proximity of their ``places of birth''.
 \citet{bla82} found an asymmetric distribution of
the light of both galaxies:  $ 38  \ \% $ of the total flux comes from the two adjacent halves of the galaxies.
According to him, this dimming of coincidental sides of binary galaxies is  observed in other interacting pairs
\citep{arp66}  and
can be due to the time scale of the interaction (not large enough to average the density of perturbed material
and subsequent star formation) and/or to obscuration of one side of one of the galaxies by the other galaxy.

NGC 5426 is a type SA(s)c pec \citep{dev91} and luminosity class
II \citep{bla82}. A heliocentric systemic velocity of $2584 \  km
\   s^{-1} $ was found through long-slit spectroscopical  observations
\citep{scha87}. From photometric observations, long-slit
spectroscopy and using  the method established by \citet{nord73},
\citet{bla82} found a mass-to-light ratio of  $(M/L)_V \ =
\ 4.1 \pm 0.6$ and $(M/L)_B \ = \ 3.9 \pm 0.7$ within a radius of
$22.5 \ kpc$.  These values include corrections for internal absorption. 
NGC 5427 is a SB(s)c pec, Seyfert 2 galaxy with  luminosity class
I \citep{dev91}. Through tertiary  indicators, it is located at
$26.7 \ Mpc$ \citep{dev79} with a  systemic heliocentric velocity
 of  $2730  \  km \  s^{-1} $ \citep{scha87}.
Up to a radius of $14.5 \ kpc$, this galaxy  has
$(M/L)_V \ = \ 2.2 \pm 0.3$ and $(M/L)_B \ = \ 1.7 \pm 0.2$ \citep{bla82}.
\citet{gonz92} find  a high rate  of star formation in the galaxy
disk, as well as a higher number of HII regions  (both giant and
supergiant)   than  expected for its  morphological  type.
\citet{alf01} obtained the rotation curve of this galaxy  through long-slit spectroscopy 
along its major axis. They  identified several oscillations along this curve that seem to be  
correlated to the presence of spiral arms and suggest taht this behavior is similar to the 
one expected in a galactic ``bore'' (or hole)  generated by the interaction of a spiral density
wave with a thick gaseous disk.
Table 2 lists the main  parameters of each galaxy.

\begin{figure*}
\centering
 \includegraphics {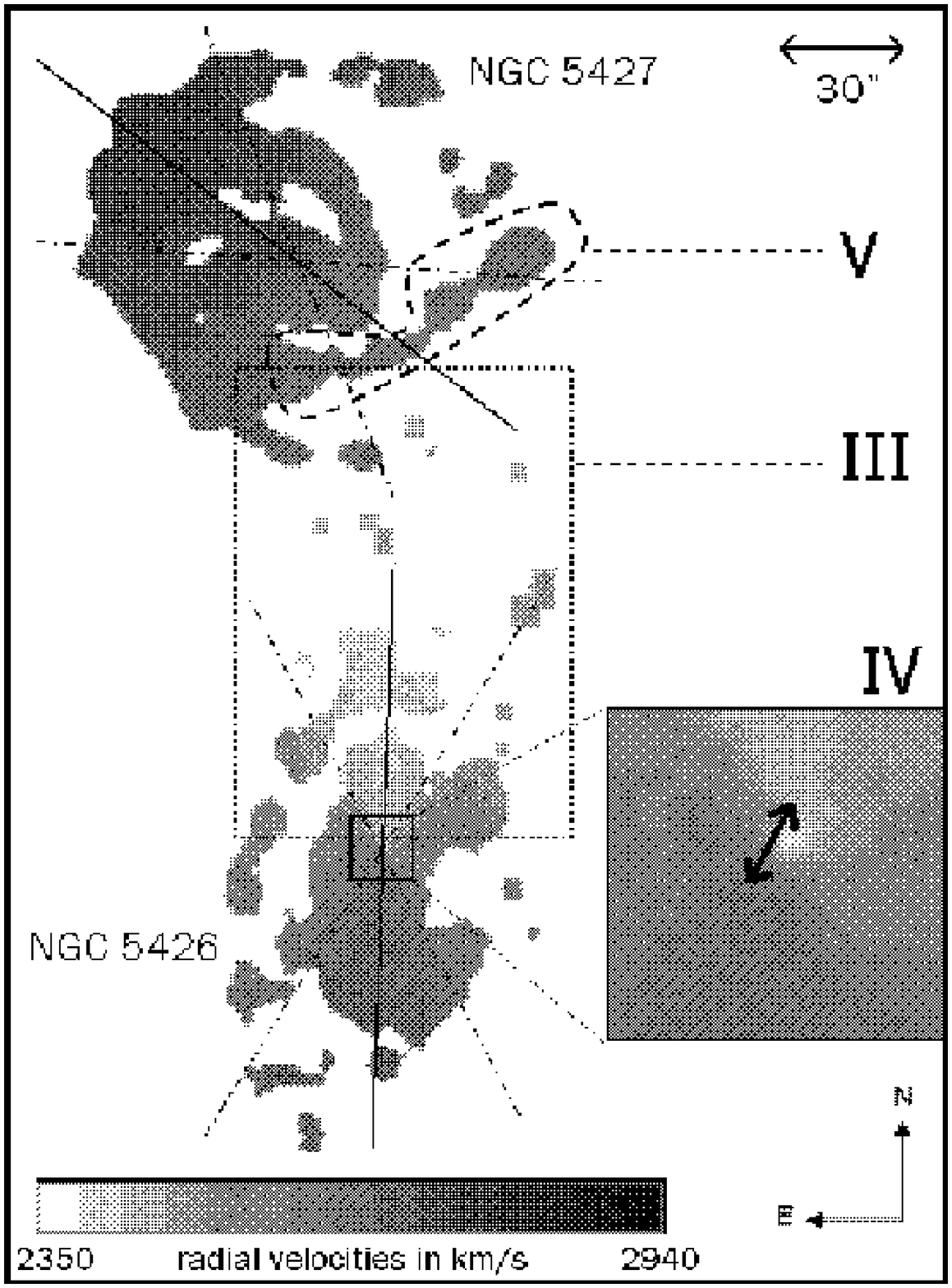}
\caption{Velocity fields  of NGC 5426 and NGC 5427.  Solid lines indicate each galaxy's 
position angle $(P.A.)$,  slash-dotted lines indicate the angular sectors from both sides of the major axis 
considered for the computation of each galaxy's rotation curve.
Region III covers the bridge-like feature between both galaxies, region IV (enlarged) is 
associated with the central parts of 
NGC 5426 where a small bar-like feature is outlined. Arrow in the enlargement  
indicates the 
location and size ($6 \arcsec$) of a small bar-like feature outlined by NGC 5426's isovelocities.  
Region V indicates the straight-arm segment in NGC 5427. }
\end{figure*}

\begin{figure*}
\centering
\includegraphics{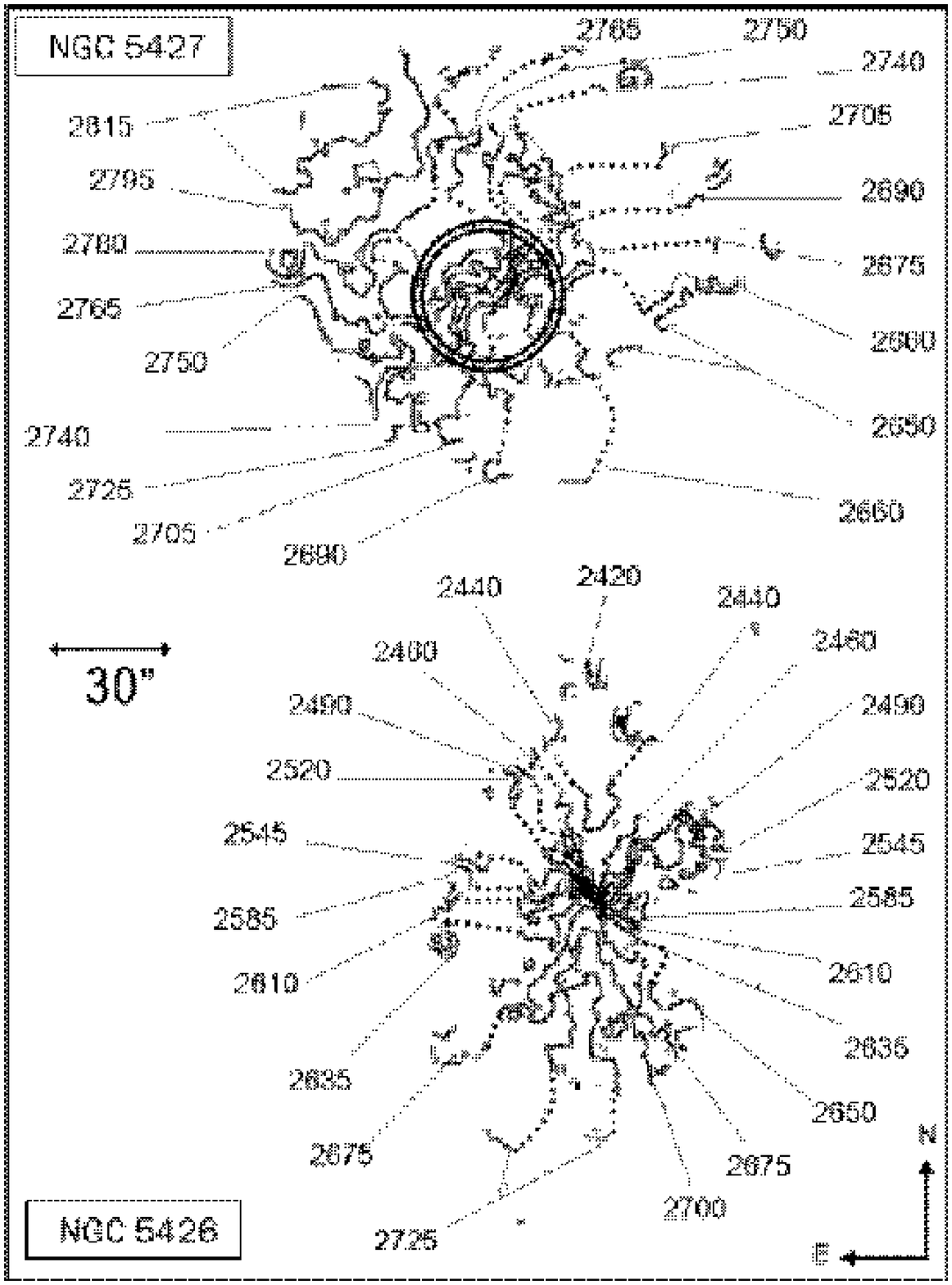}
\caption{ Isovelocities  of NGC 5426 and NGC 5427. The circled region shows central bar in NGC 5427. Velocities displayed are in $km \ s^{-1}$. Dotted lines are eye-estimated interpolation of the data.  Systemic velocities are $2575 \pm 3 \ km \ s^{-1}$ for NGC 5426 and $2722.5 \pm 1 \ km \ s^{-1}$ for NGC 5427}
\end{figure*}

\section{Velocity Fields and Rotation Curves}

\subsection{Monochromatic  Image}

Figure 1b displays the monochromatic $H\alpha$ image of the pair.
 The HII regions of NGC 5426 are  rather knotty  and they are mainly found in 
the western half of the galaxy.
The eastern and southeastern parts of the galaxy are practically
devoid of any important HII regions. 
For NGC 5427 most of the HII-regions are located 
to the north-eastern side of the galaxy along the 
western arm forming a rather patchy structure.
Many strong HII-regions appear also at the beginning
of the eastern arm, but become less intense at the
point where a straight arm segment begins. At the end of the
straight arm there is a very bright HII-region, not following
 the logarithmic spiral arm.
 The intensity of the emission decreases once the logarithmical
pattern is continued. We detect most of the HII regions detected by \citet{gonz92}.
A central bar of about $32 \arcsec $ in
total length is also detected on the monochromatic image of this galaxy.

Patches of faint monochromatic emission can be traced between both
galaxies outlining a bridge of ionized gas between both
members of the pair. This bridge follows approximately the same
orientation as the bridge of stars (continuum) that can be seen in
both the direct  image of the  pair (Figure 1a) and the
ultracontrast J-band image from \citet{bla82}. It seems to be
divided into  two  narrow monochromatic filaments:  I and II
(Figure 1b). Filament I could be interpreted as an extension of
the western arm of NGC 5426, while the location of filament II
makes it harder to associate this feature  with  any of the
galaxies' arms. 

\begin{figure*}
\centering
 \includegraphics{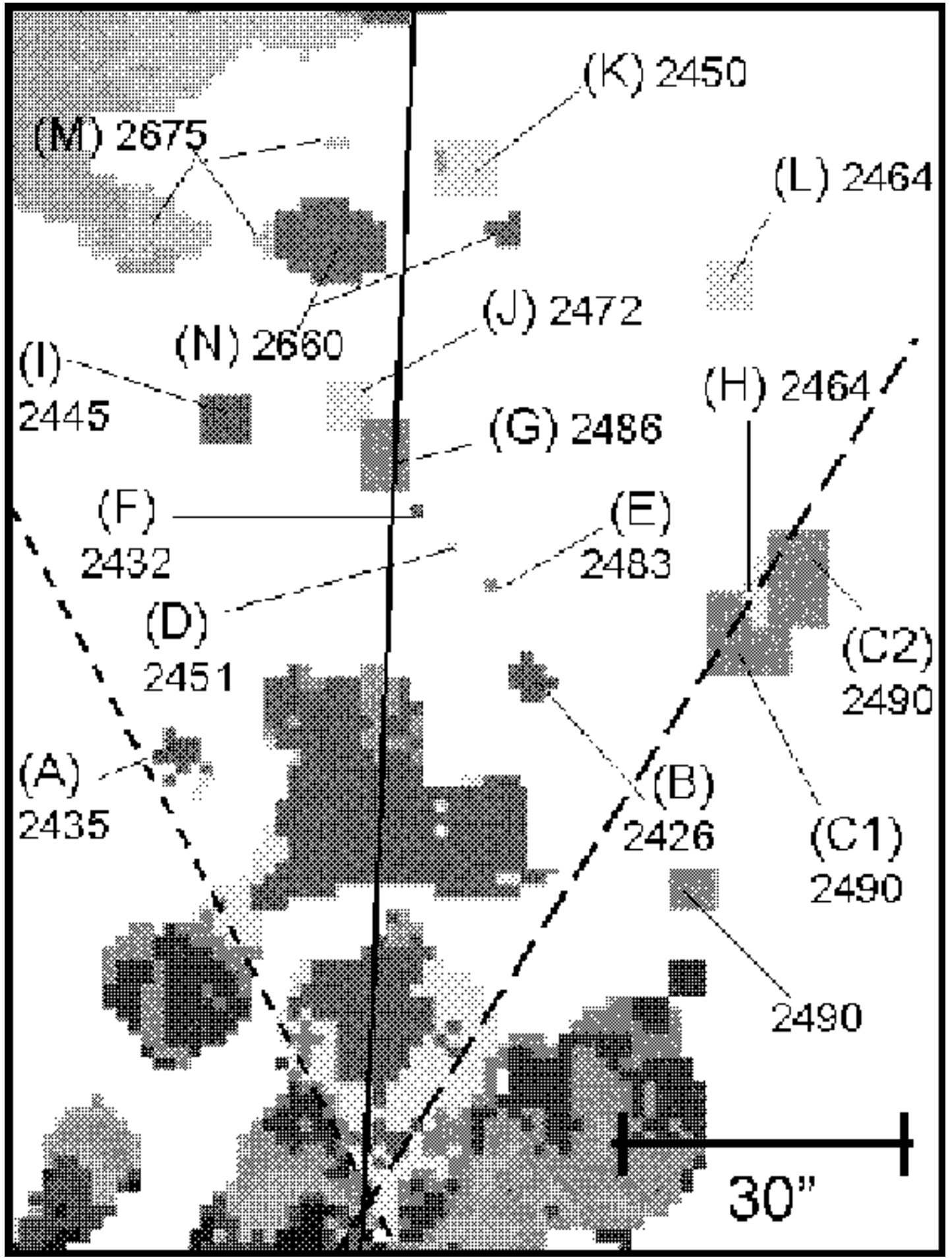}
\caption{Velocity field of the bridge-like $H_{\alpha}$ feature between NGC 5426  and NGC 5427.  
Letters are related to 
particular kinematical features  in NGC 5426's rotation curve shown in Figure 5.  }
\end{figure*}

\subsection{Velocity Fields}

Figure 2  shows the velocity fields of NGC 5426/27. Figure 3 shows the associated isovelocities.

For the main body of the galaxy, the velocity field of NGC 5426 shows little distortion if
compared  to the  velocity field of an isolated disk galaxy, though the zero-velocity line of the 
galaxy shows large wiggles.
The northern and southern parts (approaching and receding sides,  respectively) are rather symmetric.
In the outer eastern parts,  the isovelocities no longer follow the classical ``V-pattern'' of the central
parts but instead become almost horizontal. This part of the velocity field corresponds  to one of the galaxy's
 spiral arms.
In the central parts of the field (inner $6 \arcsec$), we notice  a small shift  with respect to the kinematical
center between the
 isovelocities of the receding and approaching sides  (marked with an arrow in region IV, Figure 2).
In this region, the isovelocities are almost parallel to each other following a 
southwestern-northeastern direction along a distance of about $ 6 \arcsec $ (Figure 3). This behaviour
which can be  related to a rigid-body-like  rotation, could be associated to the presence of an incipient small 
bar-like feature perpendicular to these isovelocities and of a length of about $ 6 \arcsec$.

For NGC 5427, the isovelocities on the approaching (SW) and receding (NE) sides are more opened 
and patchy
 than the ones for NGC 5426. The zero-velocity line of this galaxy also shows large wiggles 
mostly in the outer parts of the galaxy.
The isovelocities on the western side (closest to NGC 5426) show no clear orientation.
 More than one third of the straight arm has a mean radial velocity of $ (2652 \pm 7) \  km \   s^{-1}$
 while other third has a different mean value   of $( 2682 \pm 7) \  km  \  s^{-1}$  (Region V  in Figure 2).
This velocity distribution differs considerably from the
isovelocities observed on the other arm of NGC 5427  (western arm
and its bifurcation) where the  range of observed radial
velocities is much  broader. 
In the central parts rather parallel isovelocities are found within the inner $30 \pm 2  \arcsec$  
(circled region in Figure 3). These isovelocities outline the  central bar visible in the monochromatic image.

The radial velocities associated to the HII regions outlining the bridge between both galaxies
(regions I and II in Figure 1b)  are enlarged in Figure 4.
Most of these regions have radial velocities closer to those of NGC 5426  (regions A to L) 
and are furthermore suspected to belong to this galaxy.  Only regions on the northeastern 
part of the bridge have radial velocities closer
 to those of NGC 5427 (regions M and N) and probably belong to the spiral arm 
of NGC 5427.

\begin{figure*}
\centering
\includegraphics{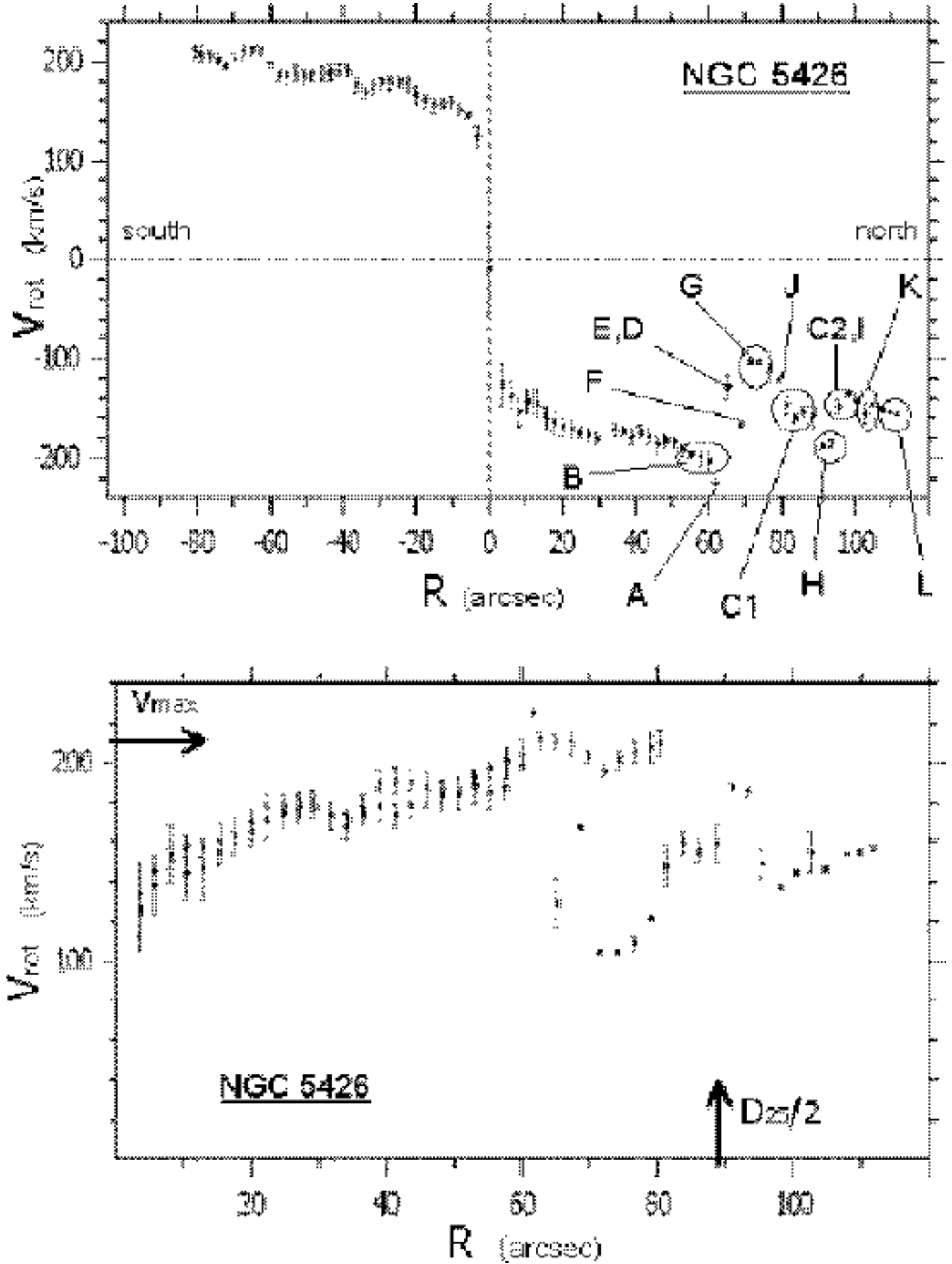}
\caption{Rotation curve  of NGC 5426. Upper panel shows the $V(R) \ vs. \ R $ for both sides of 
the galaxy -approaching and receding side. Letters are associated to the regions shown in Figure 4.
Lower panel shows the superposition of both sides. 
$V_{max}$ indicates the rotation velocity value chosen to compute the mass of this galaxy within $D_{25} / 2$
also indicated on this figure. }
\end{figure*}

\begin{figure*}
\centering
\includegraphics{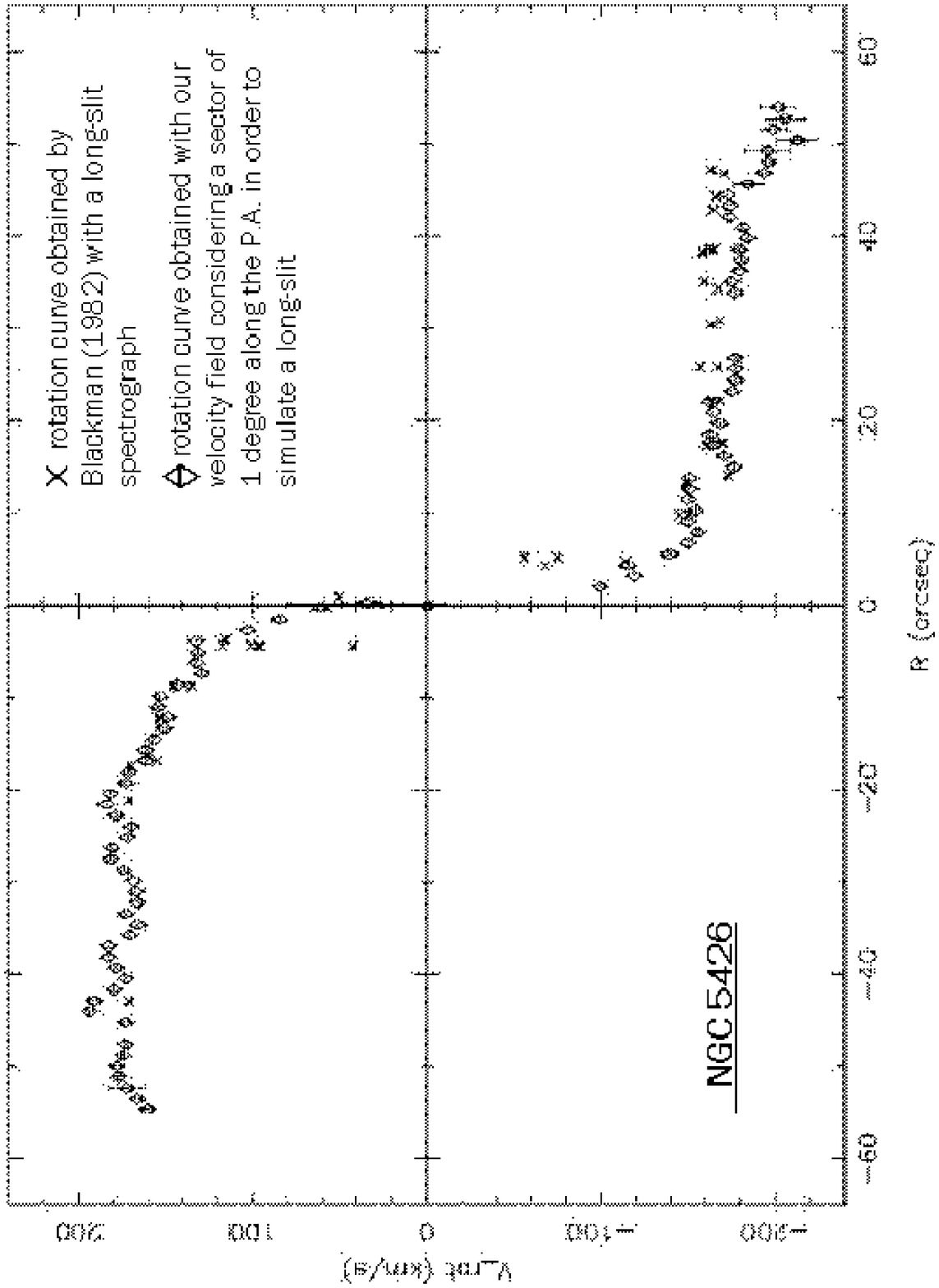}
\caption{Comparison between the rotation curve obtained by \citet{bla82} with a 
long-slit spectrograph (crosses) and our rotation curve obtained by simulating a long-slit 
spectrograph rotation curve (diamonds) considering the same parameters as that author 
($P.A$, $i$, center, angular sector)  }
\end{figure*}

\subsection{Rotation Curves}  \label{bozomath}

The interacting process perturbs each of the galaxies involved in the encounter and introduces
an important contribution of non-circular movements in the velocity fields.
Nevertheless,  in the case of early-stage  interactions, one can assume that  the inner parts of galaxies are not
strongly perturbed,  so that the velocity fields are still smooth
and symmetrical resulting in symmetric and low-scattered rotation curves up to a certain radius.

With this assumption in mind, the rotation curve of each galaxy was computed considering different
values for the kinematical parameters involved (heliocentric systemic velocity $V_{syst}$, 
kinematical center,
inclination $i$ and  position angle $P.A.$). The values were chosen such to obtain a symmetric curve in the
 inner parts of the galaxy and to minimize scatter on each side of the curve -for further detail on the 
computation of the rotation curve  see \citet{amr96}.
The first step to reduce this scatter was  to consider  points on the velocity field within a certain angular sector
on each side of the galaxy's  major axis.  The value of $i$ was then computed from the ratio between the major and minor axis and the 
$P.A.$ from the orientation of the latter. If a single ellipse was difficult to adjust to the 
continuum image,  we  considered an ellipse whose  semi-minor axis corresponded  to the kinematical 
semi-minor axis on the velocity field of the galaxy and whose semi-major axis was given by  the  
furthest emission point on the continuum image along a line perpendicular to the  
semi-minor kinematical axis.

\begin{figure*}
\centering
\includegraphics{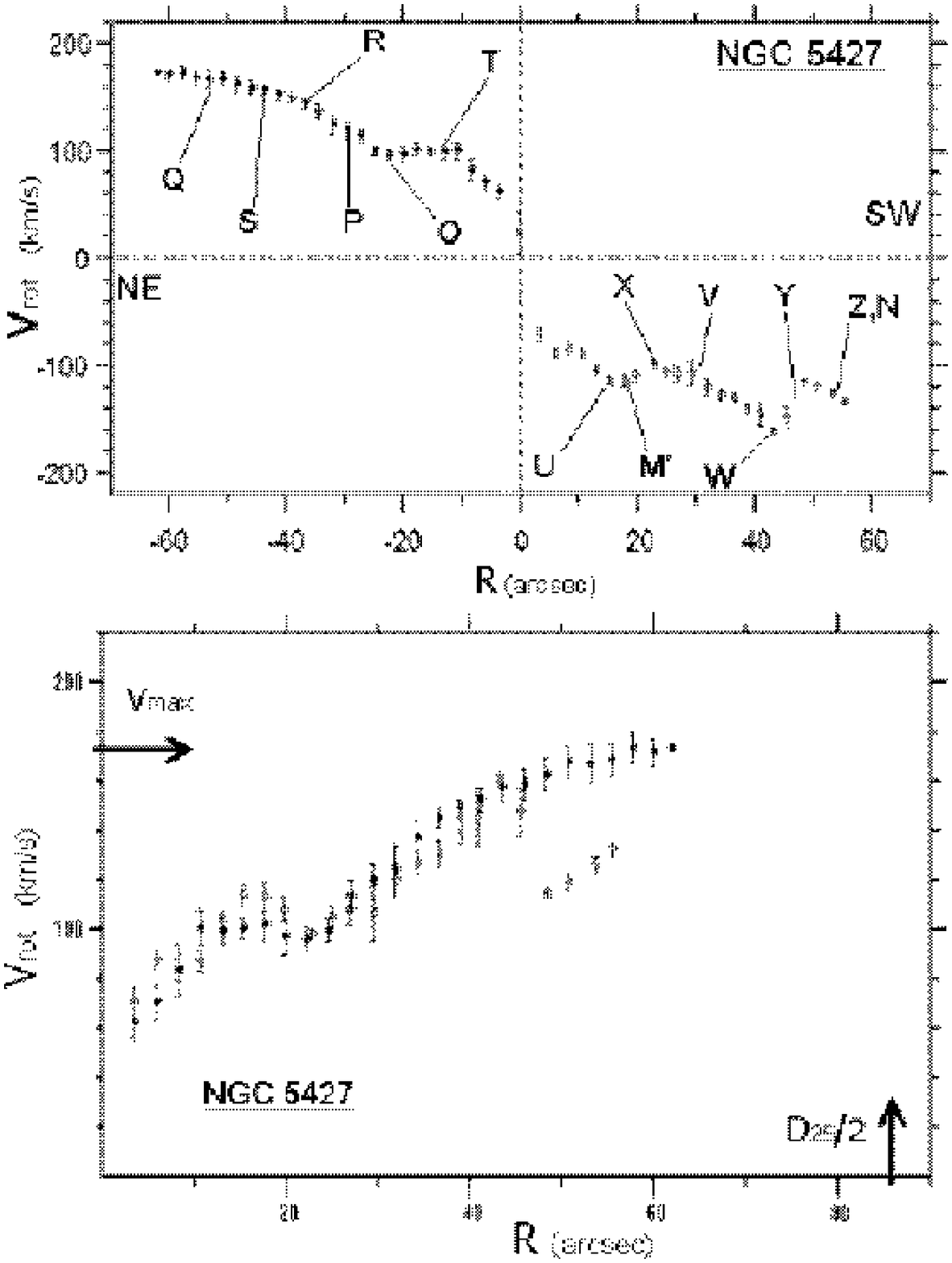}
\caption{Rotation curve  of NGC 5427. Upper panel shows the $V(R) \ vs. \ R $ for both sides of 
the galaxy -approaching and receding side. Letters are associated to the regions shown in Figure 12.
Lower panel shows the superposition of both sides. 
$V_{max}$ indicates the rotation velocity value chosen to compute the mass of this galaxy within $D_{25} / 2$
also indicated on this figure }
\end{figure*}

\begin{figure*}
\centering
\includegraphics{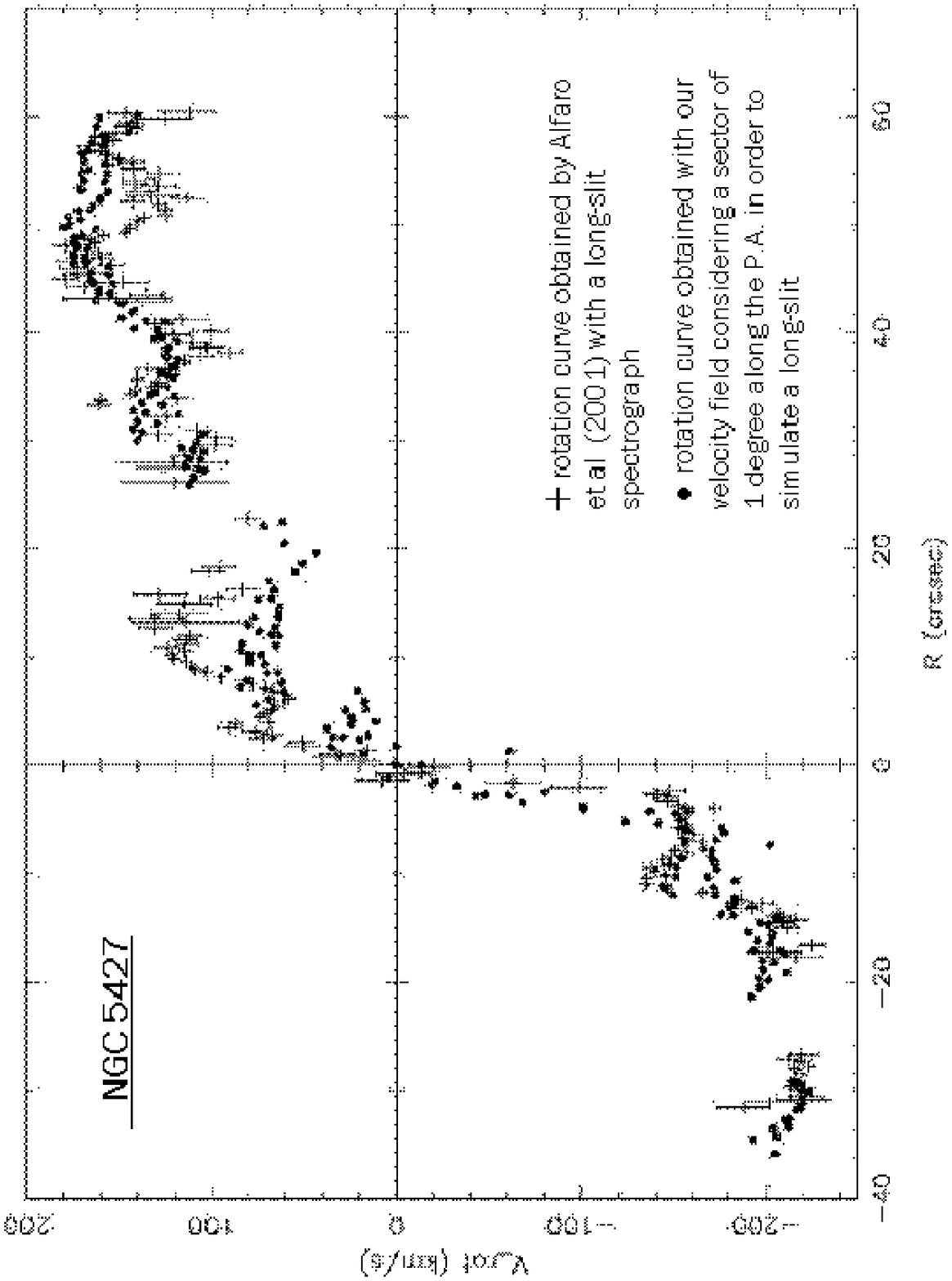}
\caption{ Comparison between the rotation curve obtained by \citet{alf01} with a 
long-slit spectrograph (crosses) and our rotation curve obtained by simulating a long-slit 
spectrograph rotation curve (dots) considering the same parameters as those authors 
($P.A$, $i$, center, angular sector)    }
\end{figure*}

\subsubsection{NGC 5426}

The rotation curve of NGC 5426 was computed considering  points on the velocity field 
within an angular sector of $ 30 ^\circ $  on each side of the galaxy's  $P.A.$
For this galaxy, the kinematical center used
to compute the rotation curve  matches the  photometric center within $ 1 \arcsec $.
 Within $ 60 \arcsec $, the set of values that resulted in the most symmetrical, smooth and less-scattered
 curve was   $P.A. =  (177.5  \pm 1)   ^\circ   $,  $i =  (59 \pm 3)   ^\circ   $ and
 $V_{sys} = (2575  \pm 3) \  km \  s^{-1} $. These are
presented in Table 2.  Figure 5b shows that  the rotation curve is rather symmetrical
in its inner $ 60 \arcsec $ .
After the  steep solid-body rotation curve at $R < 3.5 \arcsec$, it
rises slowly from $130 \ km \ s^{-1}$  to  $200 \ km \ s^{-1}$ at $R = 65 \arcsec$.
The rotation curve  shows  slight
 oscillations in phase on both sides.
 For the receding side (southern side, opposite from the companion), the rotation curve
keeps rising  with smooth oscillations up to the last detected
emission point at $ R = 80 \arcsec $. At this radius, the rotational velocity reaches its maximum value  
$ V_{max} = 209  \ km \  s^{-1}$ close to   $200  \ km \   s^{-1}$ found  with  HI observations
 \citep{bot84}.
For the approaching side (north side, towards the companion), at $R=65 \arcsec $  the rotational 
velocity abruptly falls to a value of $  (128 \pm 12)  \ km \  s^{-1} $.
Between $62 \arcsec$ and $80 \arcsec$, the rotational velocity drops to $\sim 100 \  km \  s^{-1} $.
From $ R = 82 \arcsec $, an average value of $  150  \ km \  s^{-1} $ is kept  up  to $R=110  \arcsec $.
Comparing with previous values, our value for $V_{syst}$ is similar to  the one found
by \citet{scha87}, nevertheless there is  an important difference with the value of the
$V_{syst}$  determined by \citet{bla82}.
To compare our rotation curve with the one obtained by \citet{bla82},
a long-slit  was simulated by considering radial velocities on the velocity field within 
a sector of $ 1 ^\circ $ on each side of  $P.A.$.
We then had to correct Blackman's curve by the inclination of the disk in order to superpose
both curves. The shape and amplitude of both rotation curves are very similar up 
to $50 \arcsec $ (Figure 6).

\subsubsection{NGC 5427}

An  angular sector of $ 32^\circ $  on each side of  the galaxy's  $P.A.$  was considered for
the computation of the rotation curve of NGC 5427 in order to reduce asymmetry 
and scatter up to a radius of $47 \arcsec $.
Since more than a single ellipse  could be adjusted to the 
irregular outermost isophote of the  continuum  
image, we  considered an ellipse whose
semi-minor axis corresponded to the kinematical semi-minor axis on the
velocity field of the galaxy  (radial velocities corresponding to $ \sim 2740 \  km \  s^{-1} $  
in Figure 3) to compute the  inclination $i$  and the  $P.A.$ of this galaxy.
The semi-major axis of the resulting ellipse  was then given by the furthest emission 
point on the continuum image
along a line perpendicular to the  semi-minor kinematical axis.  This perpendicular line 
was taken as the direction
of the $P.A.$
The set of values that resulted in the most symmetrical, smooth and less-scattered
 curve was   $P.A. =  (53.2 \pm 3)    ^\circ$,  $i =  (34 \pm 2.5)   ^\circ $ and   
$V_{sys} = (2722.50 \pm 1)  \ km \  s^{-1} $  (Table 2).
The kinematical center used to compute the rotation curve of NGC 5427  matches the  photometric one
within $1 \arcsec $. Globally the rotation curve displays an ascending behavior
reaching $V_{rot} = 154 \ km \  s^{-1} $ at $R = 46.5 \arcsec $
(Figure 7). There is however a trough  on both sides of the curve
between $18 \arcsec $ and $29 \arcsec $ . Up to $R = 46.5 \arcsec
$, the curve is rather symmetric. After this point, the radial
velocities on the  approaching side (south to the companion)
decrease by an abrupt shift of $50 \ km \   s^{-1} $ with respect
to the receding side. On the receding side, the curve  increases
up to $V_{rot} = 172  \ km \  s^{-1} $. From $R = 51 \arcsec $
onwards the curve flattens around a value of  $ \sim 172  \ km \
s^{-1} $ up to the last emission point on this side of the galaxy
at $ 62 \arcsec $. Our resulting rotation curve resembles the one found  by
\citet{bla82}, both in its general form as well as in the maximum
rotational velocity $\sim 175 \ km \   s^{-1}$. In order to
compare our rotation curve to the rotation curve obtained by
\citet{alf01}, a long-slit was simulated on our observed velocity
field along the same position as these authors and corrected with
their  inclination and $P.A.$ ($ 30^\circ$   and $70^\circ$,
respectively) finding the same oscillating behavior on both sides
of the galaxy (Figure 8). $V_{syst}$ by \citet{bla82},
\citet{scha87} and \citet{kee96} are in good agreement with ours. 
For the inclination $i$  and the  $P.A.$,   important differences are
found between the values determined in different works.  
Our values are larger than those found by \citet{kee96} and \citet{bla82} who also 
used kinematical considerations -see Table 2.
For NGC 5427,  the inclination determination in our case is partly  based on isophotal shape which
makes it  rather  uncertain since for inclinations of $ \sim 30^\circ$, the deviations from circular 
shape are small ($ \cos 34^\circ = 0.83,  \ \cos 24^\circ =0.91$) which are hard to distinguish. Plus 
the orientation effect on isophotal shape might easily be over-shadowed by the effects of spiral arms 
or mild deformations due to the interaction (any such effect is likely to increase the deduced inclination).

\begin{figure*}
\centering
\includegraphics{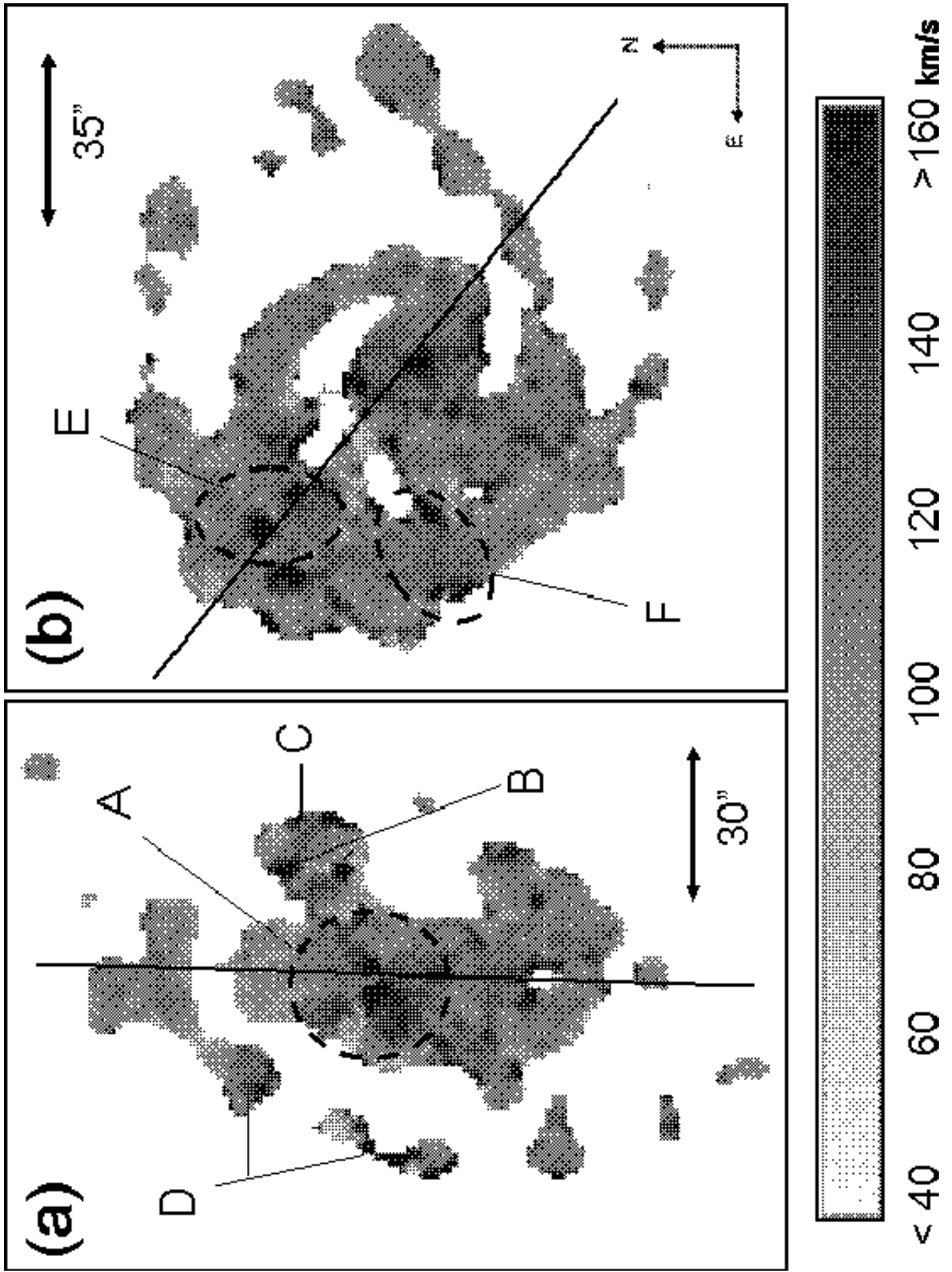}
\caption{FWHM fields  a) NGC 5426. The black ellipse shows the central cross-like 
pattern discussed in section 5.1.2.  Letters indicate regions with large FWHM.
b)NGC 5427.  Black ellipses display regions discussed 
in section 5.1.1. Regions E and F show large FWHM. }
\end{figure*}

\section{Non-circular motions}

Two dimensional kinematical fields of disk galaxies allow us to study the motion of the gas all over the galaxy and to match it with structures found through photometric methods in order to determine to which extent the gas is following pure circular motion around the center of the galaxy and to which extent there are important contributions from non-circular velocities (radial and azimuthal). 

\subsection{Full width at half maximum (FWHM)}

The FWHM of the profile at each pixel was computed from the intensity profile 
at each pixel 
as described in section 2.2
Depending on the inclination of the galaxy, this width may be associated to
motions of the gas perpendicular to the plane of the galaxy (in the case where
 the galaxy is almost face-on, 
$i \sim 0^\circ $) or to non-circular motions in the plane of the galaxy (if 
the galaxy is almost edge-on, 
$i \sim 90^\circ$). For intermediate values of $i$, the FWHM is 
associated to both types
of motion. FWHM fields for each galaxy are shown in Figure 9.

\subsubsection{NGC 5426}

The central HII regions  of this galaxy display important
values of FWHM ($ > 140  \ km \  s^{-1} $). However the
amplitude of FWHM is not simply correlated to the flux of
these HII regions. Large FWHM amplitudes ($ \sim 170 \ km \
s^{-1}$) are slightly shifted to less intense HII regions. They  match
the extremities of the small bar-like feature outlined in the
velocity field  and thus could be associated with  the motion of gas due
to the presence of  this incipient bar. A peculiar $ \sim 35
\arcsec$    cross-like pattern (NE-SW and SE-NW)   of large FWHM
 appears in the main body of the galaxy (region A in Figure 9a).
 It  is  not  associated with any particular morphological
feature in the monochromatic image  but it  does  coincide with the orientation of 
the radial isovelocities in this part of the galaxy.
Large FWHM  ($ > 140 \ km \ s^{-1}$)  are also found along the spiral 
arm of the galaxy (regions B, C and D in Figure 9a).

\subsubsection{NGC 5427}

Large FWHM values are found in the central regions of NGC 5427 matching  the
maximum of the monochromatic image and could be associated to the star-forming processes in these
regions. Nevertheless they do not follow the orientation of the central bar.
In general there seems to be a trend in NGC 5427 showing that the FWHM 
 increases
toward the faint edges of the spiral arms, both in the eastern and in the western arms, at least in those parts 
of the spiral arms where the arms are strong.
The northeastern and eastern parts of the galaxy show regions with larger FWHM
 are seen  (regions E and F  in Figure 9b).  These regions do
not coincide with any HII region, on the contrary, they lie on each side of this galaxy's western spiral arm
 where emission decreases considerably. 
In particular region E falls on the region observed by \citet{alf01} with a long-slit spectrograph and 
where oscillations on their rotation curve are interpreted as gas being
decelerated as it enters the spiral arm and  jumping above the arm to land on a  region on the
other side of the arm. 
Since NGC 5427 has a  low inclination ($ i = 34^\circ$),  the large FWHM  observed
in regions E and F can actually be associated to the perpendicular motion of the gas before and
after  it encounters the spiral arm as described by  these authors.

\begin{figure*}
\centering
\includegraphics{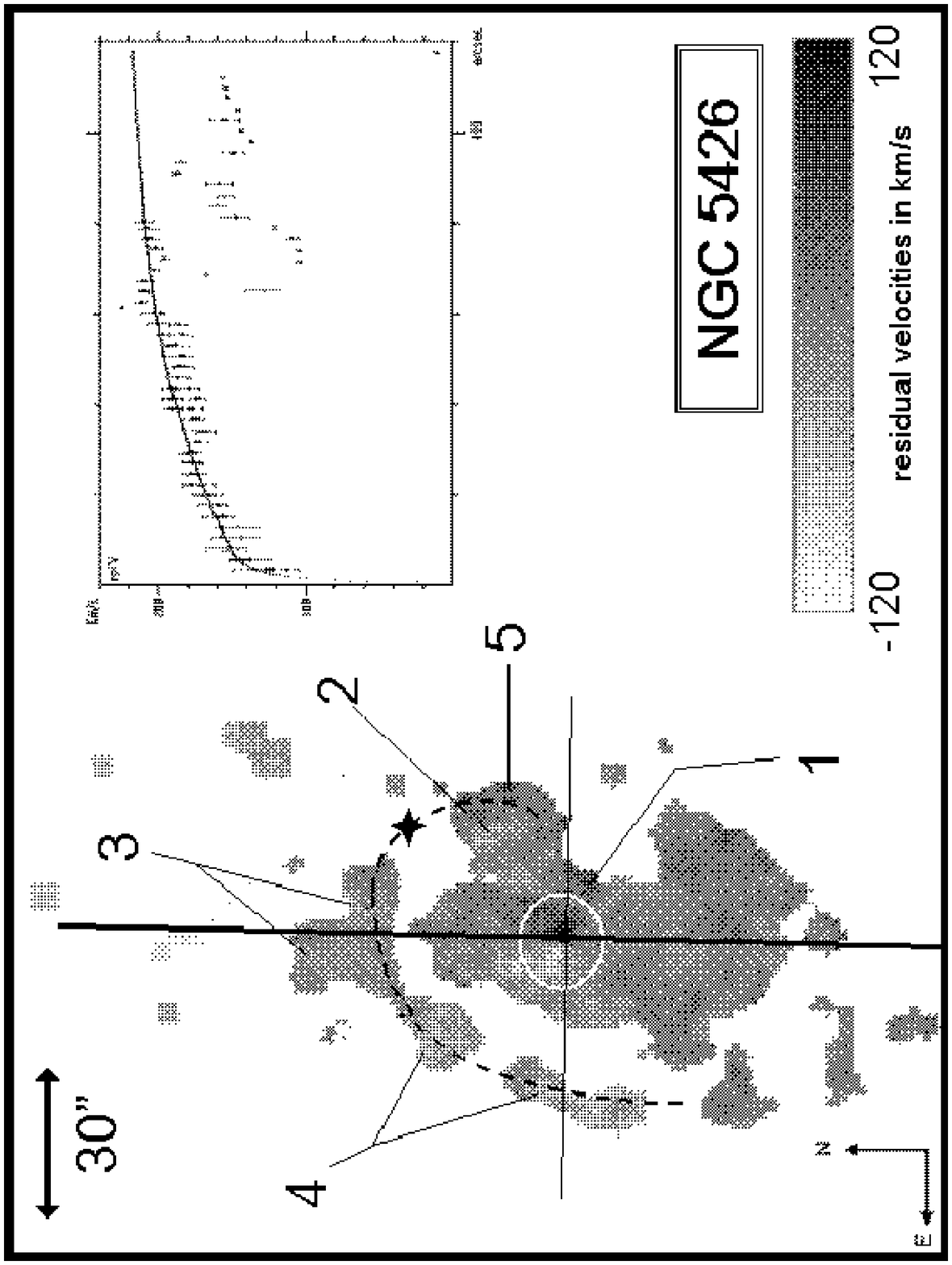}
\caption{ Residual velocity field of NGC 5426. Upper right panel shows the analytical rotation  curve 
used for the computation of
the {\it ideal} velocity field. The white circle shows the central  part of the galaxy 
where important residual velocities seem to 
indicate radial motions of the gas. Regions 1 to 5 are discussed in section 5.2.1.  The 
drawn spiral indicates location of the western spiral arm of the galaxy outlined by stars in Figure 1a. 
The black star indicates the  point at which large negative residual velocites change their location 
along the spiral arm (concave to convex) and could be related to the corotation radius of the galaxy
 }
\end{figure*}

\subsection{Residual velocity fields}

Residual velocity fields of both galaxies were obtained by
subtracting the observed radial velocity field from a {\it ideal}  radial velocity field. 
The {\it ideal}  field is the radial velocity field of an ideal galaxy in which 
there are only uniform circular motions in a plane so that 

$$ V_{ideal} (R) =   V_{syst} +  \bigl(  V_{circ} (R)  \times \cos \theta  \times \sin i   \bigr)     \eqno(1) $$

\vskip.3cm

where $V_{syst}$ is the systemic velocity of the galaxy, $V_{cir}(R)$ is the  circular velocity
component at a radius $R$,   $\theta$  is the angle along the
galaxy plane, counted from the approaching side of the nodal line and increasing in 
the direction of the galaxy rotation and
$i$ is the galaxy inclination with respect to the plane of the sky.

The {\it ideal} velocity field  is  constructed from the observed rotation curve by assigning 
to all the points of the galaxy at a certain radius, the rotational velocity from the rotation curve 
at that particular radius. These fields  prove to be very useful
to evaluate the validity of the kinematical parameters chosen to
compute the rotation curve of a disk galaxy \citep{war73}.  They
also give the opportunity to detect and analyze non-circular
motions of the gas in each galaxy since points where the residual
velocity $ V_{res} \sim 0$   correspond to regions in the galaxy
where circular motions predominate.  For points where $ V_{res}> 0 $
or $ V_{res} < 0$, the interpretation becomes more complicated.
In the general case,  the observed radial velocity $V_{obs}$ has the following
decomposition:

$$  V_{obs} =   V_{syst} + \Bigl[ \  ( V_{circ} + V_{tan} )  \times \cos \theta \ \  \ \ \ \ $$

\vskip-0.3cm

$$ \ \ \  + V_{rad} \times \sin \theta \ \Bigr]  \times \sin i   + \ \ V_{\bot}  \times \cos  i   \eqno(2) $$

\vskip.3cm

where $V_{tan}$ is the tangential velocity component in the plane of the galaxy  
(additional to the circular velocity
component $V_{cir}$ and such that $V_{rot} =  V_{cir}  +   V_{tan}$ ),  $V_{rad}$ 
is the radial velocity component in the plane
of the galaxy, and   $V_{\bot}$  is the velocity component perpendicular to the plane of 
the galaxy.

By definition the residual velocity   $V_{res}= V_{ideal} - V_{obs}$   at each point is given by

\vskip.3cm

$$  V_{res} = -  \Bigl[  \  V_{tan} \times \cos \theta + V_{rad} \times  \sin \theta  \  \Bigr]  \times \sin i  $$ 

\vskip-0.4cm

$$ \ \ \ \ \ \ \ \ \ \ \ \ \ \ \ \ \ \ \ \ \ \ \ \ \ \ \ \ \ \ \ \ \ \ \ \ \ \ \ \ \ \ \ \ \   - \ \   V_{\bot}  \times \cos  i   \eqno(3) $$

\vskip.3cm

If   $V_{\bot} \simeq 0$,  the sign of $V_{res}$ is a combination of the values of 
$V_{tan}$, $V_{rad}$ and $\theta$.
For instance along the minor axis, $\theta =90^\circ, \  270^\circ$ implies
$V_{tan}$ contribution is insignificant. 
For $\theta = 90^\circ$, $V_{rad}$  has the opposite sign of  $V_{res}$, 
while for $\theta = 270^\circ$,   $V_{rad}$  has the same sign as $V_{res}$.
Along the major axis $V_{rad}$ contribution is insignificant  so that  
for $\theta = 180^\circ$,   $V_{tan}$  has the same sign
as  $V_{res}$, while for $\theta = 0^\circ$, $V_{tan}$  has the opposite sign of  $V_{res}$.

\begin{figure*}
\centering
\includegraphics{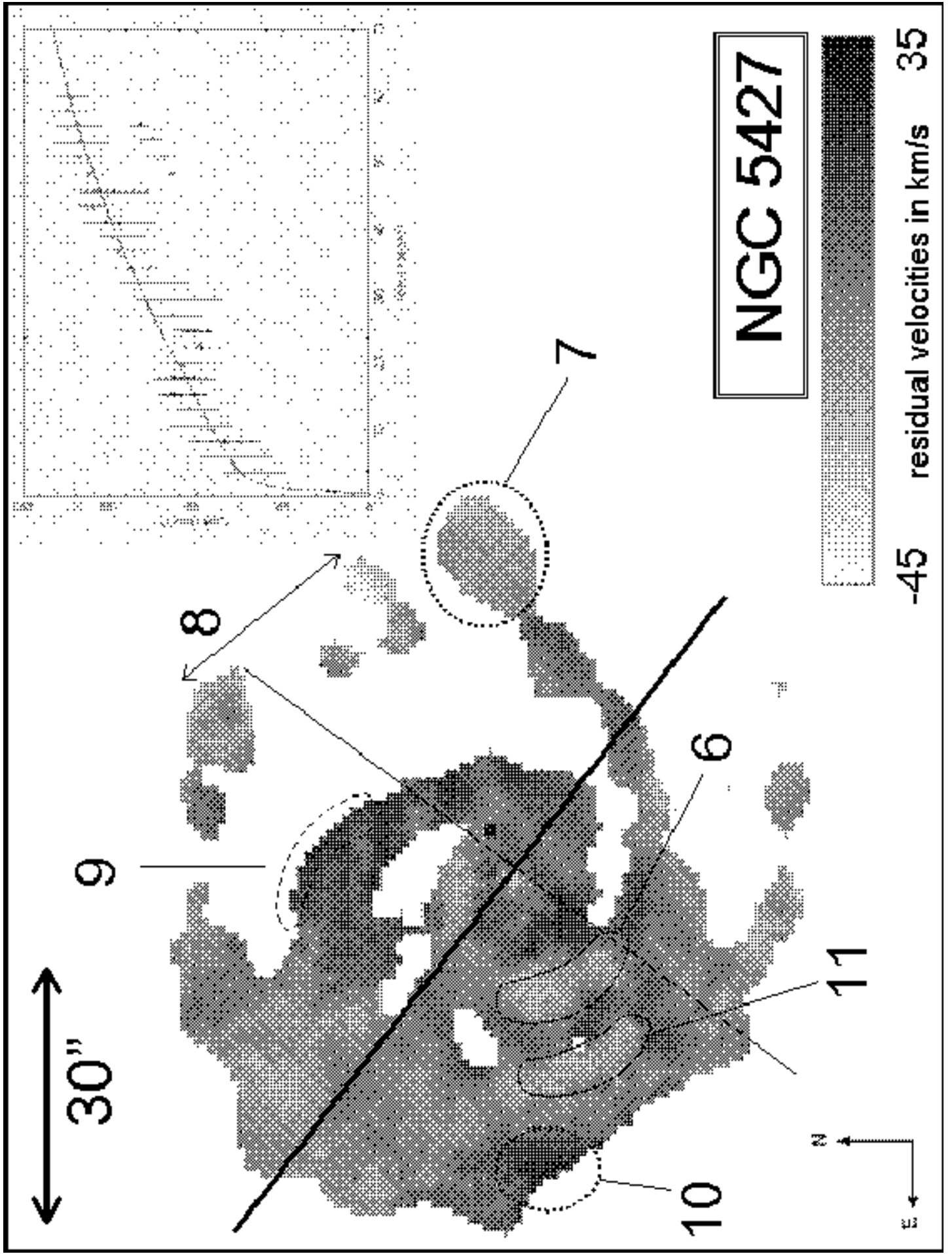}
\caption{Residual velocity field of NGC 5427. Upper right panel shows the analytical rotation curve 
used for the computation of the {\it ideal} velocity field. 
Regions 6 to 11 are discussed in section 5.2.2.   }
\end{figure*}

The analysis of non-circular motions from the residual velocity field is  complemented
 with information from the velocity dispersion field. 
For instance for low inclinations,  $V_{\bot}$   can be neglected in 
equation 2 for regions displaying small  velocity dispersion values.

\subsubsection{NGC 5426}

In order to compute the residual velocity field for this galaxy,  an analytical curve of the form
$ V(r) \propto r^{-1} \times [ r^{1/2} + exp(r) ] $ was adjusted to the observed
rotation curve omitting the points that seemed to be associated with the bridge-like feature. 
The fit is shown in the upper-right frame in Figure 10.
The {\it ideal} velocity field considered was  obtained from this rotation curve and equation 1.
Figure 10 shows  the residual velocity field of NGC 5426.
This field shows no feature associated with
an error in the determination of the kinematical parameters (see discussion in \citet{war73})
 so that the different velocities displayed
in the galaxy correspond to actual non-circular motions of the gas.
In the central part (inner $ 6 \arcsec $),  there is an important
velocity difference. The residual velocities go  from $ - 40 \ km
\ s^{-1} $ in the inner  $ 3 \arcsec $ east to the center of the
galaxy to  $ + 90 \ km  \ s^{-1} $ in the inner $ 3.5  \arcsec $
west of the center (region 1 in Figure 10). This region matches the incipient bar-like feature
detected on the velocity field.
An important contribution of negative residual velocities is also
seen along the western spiral arm. At the beginning of the spiral
arm (region 2 in Figure 10), this contribution is seen on the
concave side of the arm. On the convex side of the same arm,
positive residual  velocities are found (region 5). Further along the
arm, large negative residual velocities are seen on the convex
side up to the tip of the arm (regions 3 and 4). This peculiarity
could be explained by the acceleration of the gas as it goes
through the spiral arm as suggested by \citet{pism86} or to
vertical motions also associated to the passage of the gas through
the density wave   \citep{alf01}. This contribution of residual
velocities could be associated to both types of motion. The point
where the position of negative velocities changes from concave to
convex could be associated to the sense of passage of the gas with respect 
to the pattern speed of the spiral arms and thus could be associated to
the corotation radius of the galaxy (at $R = 50
\arcsec $). This point  is marked with a black star in Figure 10.

A comparison between the FWHM  field and the residual velocity field
shows  a correlation between  FWHM  and $V_{res}$ for the central parts of this
galaxy (Figures 9 and 10) implying an important contribution of non-circular
motions. Due to the  inclination of the galaxy
($i = 59^\circ $) we cannot  explain the large values of the
FWHM  in the central regions of this galaxy solely by the  presence of large
perpendicular motions since they could also  be the result of
important non-circular motions in the plane of the galaxy.
Considering that in these regions ($\theta \sim 90, 270 ^\circ$)  
there is little contribution of $V_{tan}$, a fraction of these non-circular motions is probably 
in the radial direction. Since this region corresponds to the small bar-like
feature outlined by the observed velocity field, these non-circular
motions could imply gas moving along this structure.

\subsubsection{NGC 5427}

For NGC 5427, the theoretical velocity field was  also derived
from  an analytical curve of the form $ V(r) \propto r^{-1} \times[ r^{1/2} + exp(r) ] $ 
fitted to the observed rotation curve omitting the points  beyond the  bifurcation radius 
with a low rotation velocity (that is points with $ V_{rot} < 115 \ km  \ s^{-1} $  
beyond $ R = 46 \arcsec$ ).  
The fit is shown in  the upper-right  frame in Figure 11.
 
Regions with $ V_{res}$ different from zero seem to be associated with the spiral arms of the galaxy.
Positive residual velocities $ V_{res} >  15  \ km  \ s^{-1} $ are
found in the region between the nucleus of the galaxy and the
beginning of the eastern arm (region 6) and the beginning of the
western arm (region 11 in Figure 11).
Positive residual velocities larger than $  15  \ km  \ s^{-1} $
are also seen along the eastern arm of the galaxy (region 8), including
the tip of the straight arm segment (region 7). Negative residual
velocities are also seen on the outer edges of the western arm (regions 9 and 10).
These residual velocities could be associated with the motion of
the gas as it encounters, goes through and leaves the density wave
as suggested by \citet{pism86}. As mentioned by \citet{alf01},
vertical motions should also be present in both cases thus
contributing to the non-circular motions at the edges of this
spiral arm. None of the regions with large $ \vert  V_{res} \vert$ 
are located along the major or minor axis of the galaxies, so no
further analysis can be done using equation 3.

\subsection{Kinematical vs Morphological Features}

\subsubsection{NGC 5426}

The bridge-like feature between both galaxies was analyzed  by
comparing  the location in the projected plane of each emitting
region  with  its associated radial velocity. Only regions M and N
(Figure 4) display radial velocities closer to those of NGC 5427
and could actually belong to this galaxy's disk. Other regions that
seem to be closer to NGC 5427 than to NGC 5426 -in projection- display velocities
closer to those of  NGC 5426  (regions G to L). We analyzed the
influence  of each of these regions on NGC 5426's rotation curve
(Figures 4 and 5) in an attempt to tell which of these regions
still follow the main motion of the galaxy's rotation and which
regions seem to be suffering the effects of tidal forces due to
the interaction.
Regions A and B still follow the global behavior of the rotation
curve, region F shows a decrease of $ 20 \ km  \ s^{-1} $ in the
rotation velocity  while adjacent regions D, E, G and  J are
clearly below the average rotation velocity of the galaxy 
($V_{rot} \sim 110 \ km  \ s^{-1}$)  compared to an average value of
$\sim 200 \ km  \ s^{-1}$ in parts of the galaxy which
seem to display a circular motion. At larger radii, regions C1, C2, I, K and L show
a slight increase of $ \sim 30 \ km  \ s^{-1}$ but they still
remain below the average rotation velocity. Only region H shows a
velocity value closer to the average. In the projected plane,
regions A and B do seem closer to NGC 5426 and could actually
belong to this galaxy's disk. Nevertheless regions E, D, G and J
seem to be closer to NGC 5426  than regions K and L whose
velocities are closer to those of the average rotation curve. This
could give a clue as to the actual position of the bridge-like
feature in a 3D scenario.
Regarding the inner structure of the galaxy, kinematical
information from the rotation curve is not available for the inner
$7 \arcsec$ of this galaxy, so any eventual effect of the small
bar-like feature  ($  \sim 3.0 \arcsec $ in semi-length)  on the
rotation curve is not visible.

\subsubsection{NGC 5427}

For the approaching side of this galaxy, most of the oscillations
observed can be associated to the HII regions along the arms of
the galaxy (Figures 7 and 12). For instance, the small bump around
a radius of $16 \arcsec $  corresponds  to the beginning of the
western arm and is marked by a small HII region (region $U$ in Figure 12). This
radius matches the semi-length of the central bar. 
The velocity then decreases by $\sim 10 \ km  \ s^{-1}$ which corresponds to
the interarm region marked as $X$. After this region,  the
rotational velocity begins to increase rather steadily once we
reach the straight segment of the eastern arm of the galaxy
(region between $V$ and $W$). We then notice an abrupt decrease in
the velocity of about $40 \ km  \ s^{-1}$ (region $Y$) between the
second-to-last HII region ($W$) and the last HII region of the straight
spiral arm segment (region $Z$)  as well as  the tip of the western arm (regions $N$). 
The tip of this  arm is marked by a very intense and
large HII region where important star formation is going on
(region $Z$). This side of NGC 5427 is closer to the
companion, so this decrease of the rotational velocity  could be due to   the slowing-down 
and homogenization of velocities along the
straight arm and/or to the fact that this part of the arm no longer lies in the
main plane of the galaxy-both these processes being caused by the interaction. 
On the receding side of the rotation curve, the  end of the bar
also matches a small bump feature (region $T$). The rotational
velocity decreases at the interarm region ($O$). Following this
decrease, the velocity increases forming a slight hump from the
bifurcation of the western arm ($P$) -marked by an  HII region- to
larger and  more intense HII regions along this arm (regions $R$
and $S$). This hump then reaches a slight plateau once 
the outermost arm segment of the western arm ($Q$) is reached.
At $R = 16 \arcsec$ on  both sides of the rotation curve, 
the end of the bar is marked by a  small plateau
of  $\sim 110 \  km  s^{-1}$ for the receding side (region  $T$) and of 
$\sim 100 \  km  \ s^{-1}$  for the approaching one (region $U$).

\begin{figure*}
\centering
\includegraphics{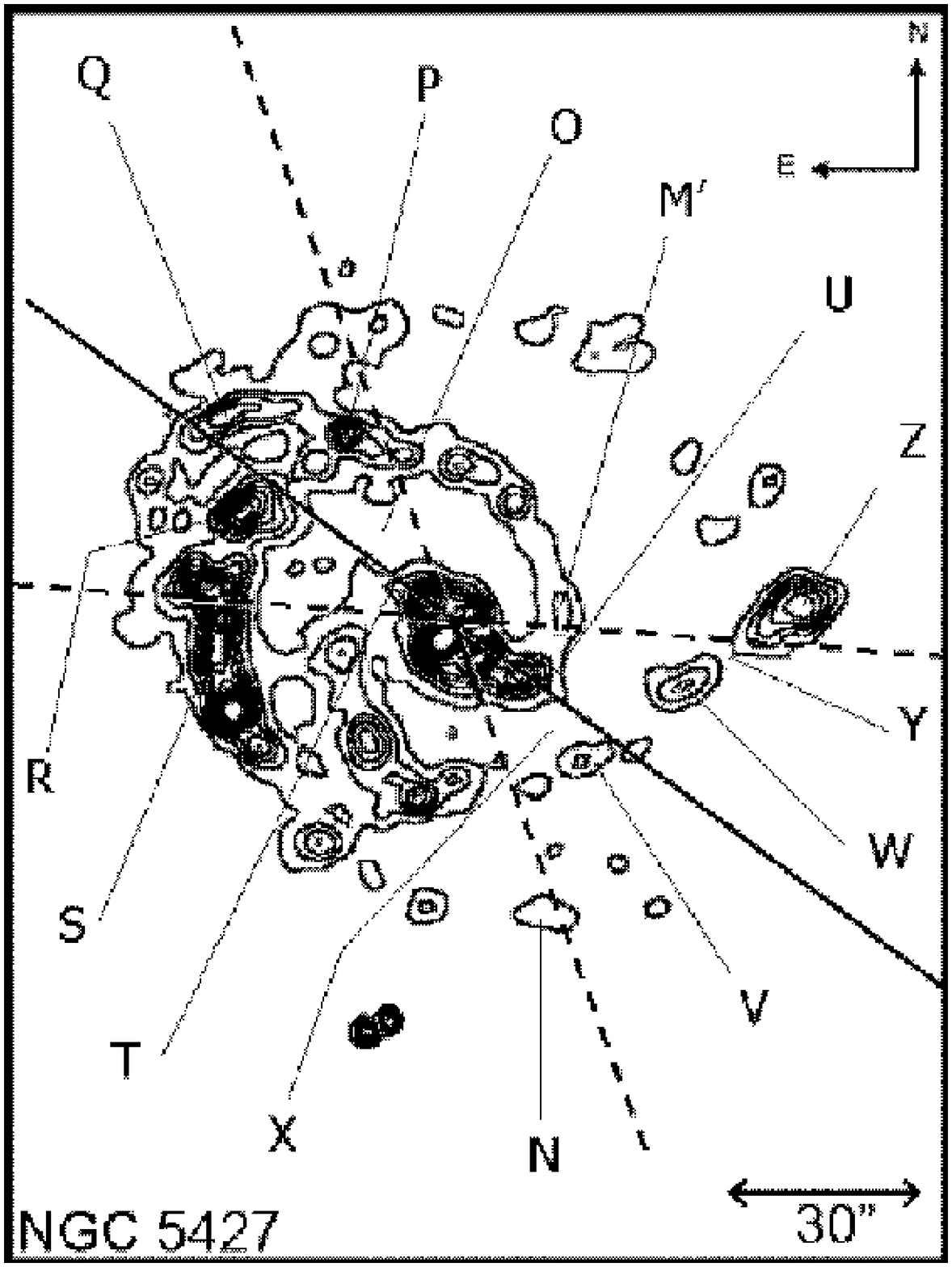}
\caption{Monochromatic isophotes of NGC 5427. Letters indicate features associated to variations in this
galaxy's rotation curve shown in Figure 7. Solid line indicates  the galaxy's 
position angle $(P.A.)$, the  slash-dotted lines indicate the angular sector from both sides of the major axis 
considered for the computation of the galaxy's rotation curve. }
\end{figure*}

\section{Dynamical Analysis}

\subsection{Mass Estimates}

For each galaxy, a range of possible masses was computed  using the method proposed  by \citet{leq83}.
This method considers two extreme cases to evaluate a galaxy's mass: the galaxy viewed
as a flat disk and the galaxy viewed as a spherical system.
For the flat disk case,   the formulas from the
mass model  by \citet{nord73} and a  theoretical rotation curve that stays flat after the
solid-body rotation section  are used as assumptions to approximate the mass inside a radius $R$  by
$ M(R) = 0.6  \times  (RV^2(R) / G) $ . On the other side for any  spherical model, the
 mass inside a certain radius is always given by  $ M(R) = RV^2(R) / G $.
For a real disk galaxy, the actual value of $M(R)$ should always lie between these two extreme  values. 
In fact, if a massive halo is considered, the expression considered for the spherical 
model  should be the most appropriate for the determination of a  galaxy's mass.

The  overall rotation  curves of these galaxies cannot be considered as flat.
Nevertheless since this is an early interaction,  circular motions of the gas dominate  the 
inner parts of the galaxy so that up to a certain radius  the rotation
curve for both sides of the galaxy is representative of circular motion in which case  
Lequeux's method may be applied.
 In order to determine this radius,  we looked  for the radius on the rotation curve,  $R_{bif}$   
at which symmetry of the approaching and the receding sides is dramatically lost
($ \vert V_{receding} - V_{approaching} \vert  \geq 40 \ km  \ s^{-1}$).

For NGC 5426 (Figure 5),  symmetry is lost after  $  R_{bif}=  60 \arcsec$,  the velocity at this 
radius is $198 \ km  \ s^{-1}$.
Considering  the distance modulus given by \citet{dev79} for NGC 5427 is also valid for this galaxy, 
 then  $R_{bif}=  7.8 \ kpc$ so that up to this
radius this galaxy's mass  lies  between
 $4.05 \times 10^{10} M_\odot$ and  $6.75 \times 10^{10} M_\odot$.
For NGC 5427 (Figure 7), $R_{bif}=  46.5  \arcsec$ with $V = 154 \ km  \ s^{-1}$. Considering the same
 distance modulus,
$R_{bif}=  6.02 \ kpc$, so the mass up this radius lies between $1.92 \times 10^{10} M_\odot$ and
$3.20 \times 10^{10} M_\odot$. Results are shown in Table 3.

Furthermore,  although for both rotation curves the side of the curve  closest  to the  companion
 shows dramatic variations after $R_{bif}$, the  other side of each curve shows  no distinct 
difference before and after this bifurcation radius is reached.  For both galaxies, the rotation curve 
seems to flatten down at large radii after reaching a maximum velocity $V_{max}$ (Figures 5
and 7). One may then assume that for each of these sides, the
rotation curve is representative of circular motion up to the last
detected emission point and that it can be extrapolated up to
$D_{25}/ 2$  by assigning to this radius the maximum rotational
velocity observed $V_{max}$.
For NGC 5426, $V_{max} = 208 \  km  \ s^{-1}$    and  
$D_{25}/ 2 = 88.6 \arcsec = 11.47 \ kpc $, 
for NGC 5427,   $V_{max} = 172 \  km  \ s^{-1}$ and
$D_{25} / 2 = 85.5 \arcsec = 11.06 \ kpc $ -both $D_{25}/ 2$  values were taken from the 
LEDA database.  
The resulting individual masses computed with these values are 
$M_{NGC 5426} = 1.12 \times  10^{11}  \   M_\odot$  and 
$M_{NGC 5427} =  0.75 \times  10^{11}  \   M_\odot$
Results are summarized in Table 3.

The combined mass of the components from the orbital motion of the pair was also computed 
 using the  expression given by \citet{kar84} for an statistical estimate of the orbital mass:

$$ M_{orbital}  = {32 \over {3 \pi }}   \   \biggl(  {  {\Delta V ^{2} \times X_{12}} \over G }  \biggr)   \eqno(4)$$

 where
$\Delta V$   is the difference between the systemic velocities of the galaxies,
$ X_{12}$ is the projected separation from one nucleus to the other,
$ G$ is the gravitational constant and
$32/3\pi$   is the mean value of the projection factor for circular motions of the members of the pair
and isotropic orientation of the orbits.
For NGC 5426 and NGC 5427, we have   $\Delta V = 147.5  \ km  \ s^{-1}$ and $X_{12} = 18.17 \ kpc$,  so  the
orbital mass equals  $ 3.01 \times  10^{11}  M_\odot$.

Comparing the orbital mass with the sum of the individual masses we find that
$M_{add} = M_{NGC 5426} +  M_{NGC 5427} =  1.87 \times  10^{11}  \   M_\odot$, so that
$  M_{add} /  M_{orbital} = 0.62$. It should be kept in mind though that equation (4) is a statistical estimate 
of the orbital mass of the pair under certain assumptions and such that  the lower mass limit for a circular 
orbit is about $1/3$ of this value and for a parabolic orbit the value is even smaller.

\begin{flushleft}
\begin{table*}
\centering
     \caption[]{Mass estimates   for NGC 5426 and NGC 5427 }
 \begin{tabular*}{1\linewidth} [  ] {l c c c c c c c c c c c c }
\hline
 \noalign{\smallskip}
              & &  $R_{bif}^{\mathrm{a}}$  &  & $V(R_{bif}) $ &  & $M(R_{bif}) \ \times 10^{10} \ M_\odot$   & -  & $D_{25}/2 $  &  & $V(D_{25}/2)$  &  & $M(D_{25} / 2) \ \times 10^{10} \ M_\odot$    \\
              &  & in $ kpc$ &  & in $km \ s^{-1}$ & &  &  &  in $ kpc$ &  &  in $km \ s^{-1}$ & &    \\
\noalign{\smallskip}
\hline
 \noalign{\smallskip}
NGC 5426   &   & $7.8 $  & & $198 $ & & $4.05 - 6.75$   & -  &  $11.5 $  & & $208 $ & & $6.72 - 11.2$   \\
  \noalign{\smallskip}   

NGC 5427  &  & $6.0 $  & & $154 $ & & $1.92 - 3.20$ & -  & $11.1 $  & &  $172 $ & &  $4.5 - 7.5$  \\

\noalign{\smallskip}
\hline
\end{tabular*}
  
\begin{list}{}{}
\item[$^{\mathrm{a}}$] Radius at which symmetry on the rotation curve (approaching vs receding side) is lost.
\end{list}

\end{table*}
\end{flushleft}

\subsection{Mass Distribution}

The mass model used in this work  was taken from  \citet{blais01}.
It  considers both the light distribution of the stars and a theoretical dark halo profile
 to compute a rotation curve that best fits the observed one. 
The density of the theoretical dark matter halo is given by the following expression \citep{blais01}:

$$ \rho (R) = { \rho_0 \over [ c + (R/R_0)^\gamma] [ 1 + (R/R_0)^\alpha]^{( \beta - \gamma) / \alpha} }  \eqno(5) $$
where $\rho_0$ is a  characteristic density, $R_0$ is a characteristic radius and $c$ can force the presence of a flat-density core. The parameters $\alpha$, $\beta$ and $\gamma$ determine the shape of the density profile.
For a pseudo-isothermal sphere \citep{beg87}, $c = 1$, $\alpha = 1$, $\beta =2 $ and $\gamma = 2$; for a NFW profile  \citep{nav96}, $c = 0$, $\alpha = 1$, $\beta =3 $ and $\gamma = 1$ and  for an isothermal sphere \citep{krav98},  $c = 0$, $\alpha = 2$, $\beta =3$ and $\gamma = 0$.
In this way, the $M/L$ as well as the dark matter  halo properties: central density $\rho_{0}$ and core radius $R_{0}$, are free parameters.
For both galaxies optical photometry in the B band was taken from \citet{woz95}.

For NGC 5426,  none of the models considered (maximum disk, submaximum disk or best fit model; 
pseudoisothermal,  NFW or isothermal dark halo profile)
 fits  the three inner points of the observed rotation curve. Figure 13 shows the fit with the 
lowest computed $\chi^2$, it considers a maximum disk ($M/L = 11.15$) and a pseudo-isothermal 
sphere  for the dark halo.
The inner points of the rotation curve could only be fitted by assuming the contribution of the disk is 
almost negligeable ($M/L = 0.1$) and considering a pseudo-isothermal halo with a large central 
density, $\rho_0 = 0.005 \ M_\odot \ pc^{-3}$ .
One possible, disk-related, reason why the inner rotation curve of NGC 5426
cannot be fitted is that  the  disk contribution considered for this fit  was taken from a B-image while 
the central regions of the
galaxy might have larger $M/L$ than the rest of the disk (the bulge region is probably
'redder').

For NGC 5427,  the best fit model gives  $(M/L)_B = 1.7 $ (maximal disk) -which 
matches the value found by \citet{bla82}-
 and  an isothermal halo  implying an important dark matter
contribution in the outer parts of this galaxy. The divergence
between the data and the model correspond to regions associated to
the spiral arms  (Figure 14).

\begin{figure*}
\centering
\includegraphics{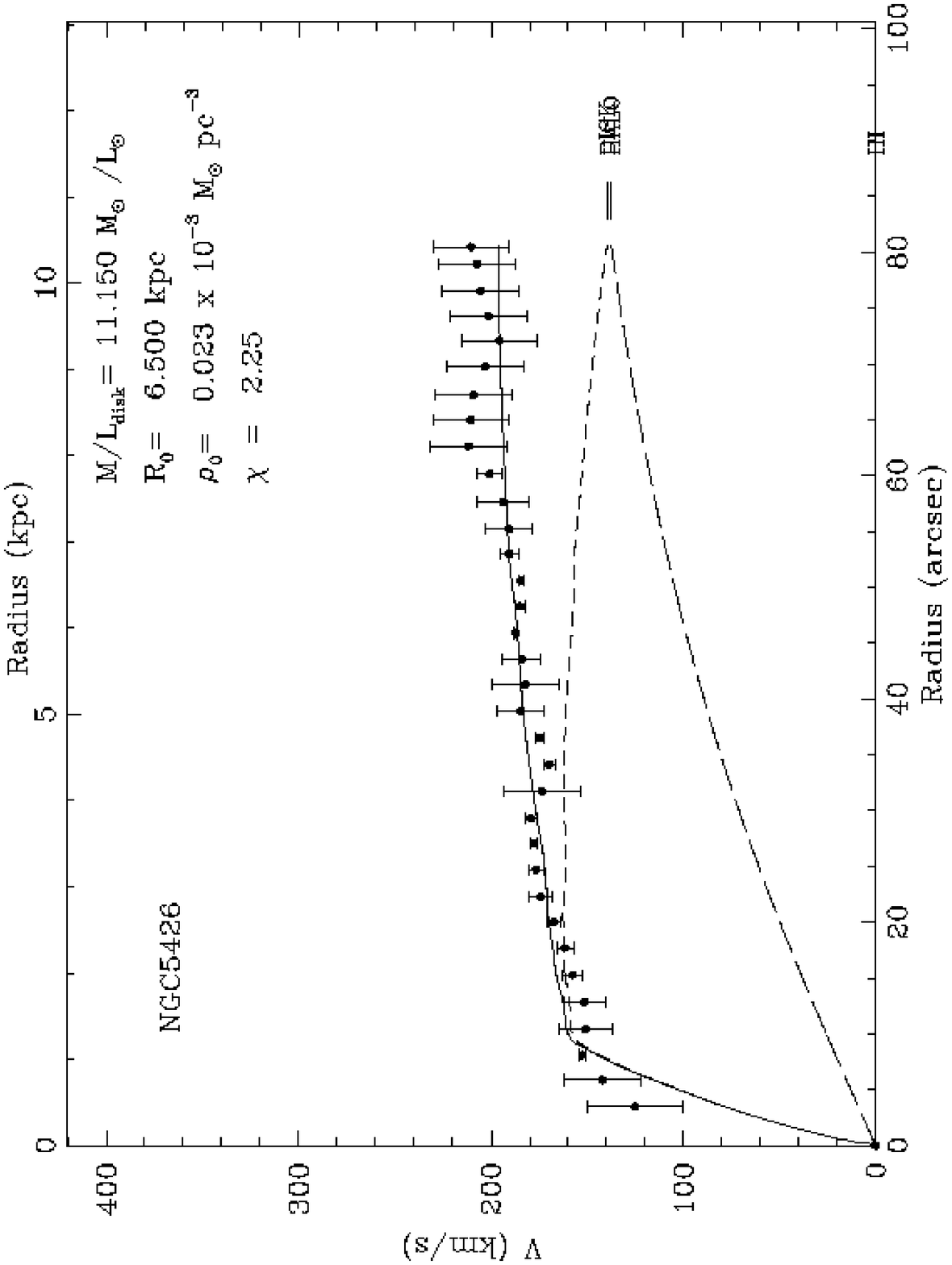}
\caption{Best mass model fit for NGC 5426 using a pseudoisothermal halo and a maximum disk. Long-dashed curve represents the dark-matter halo contribution, short-dashed curve represents the stellar disk contribution. The parameters displayed in each curve stand for the mass-to-light ration of the stellar disk ($M/L_{disk}$), the core radius of the dark matter halo and its central density ($R_0$ and $\rho_0$,  respectively) and the minimized $\chi^2$  in the three-dimensional parameter space. Mass-model taken from \citet{blais01}.}
\end{figure*}

\begin{figure*}
\centering
\includegraphics{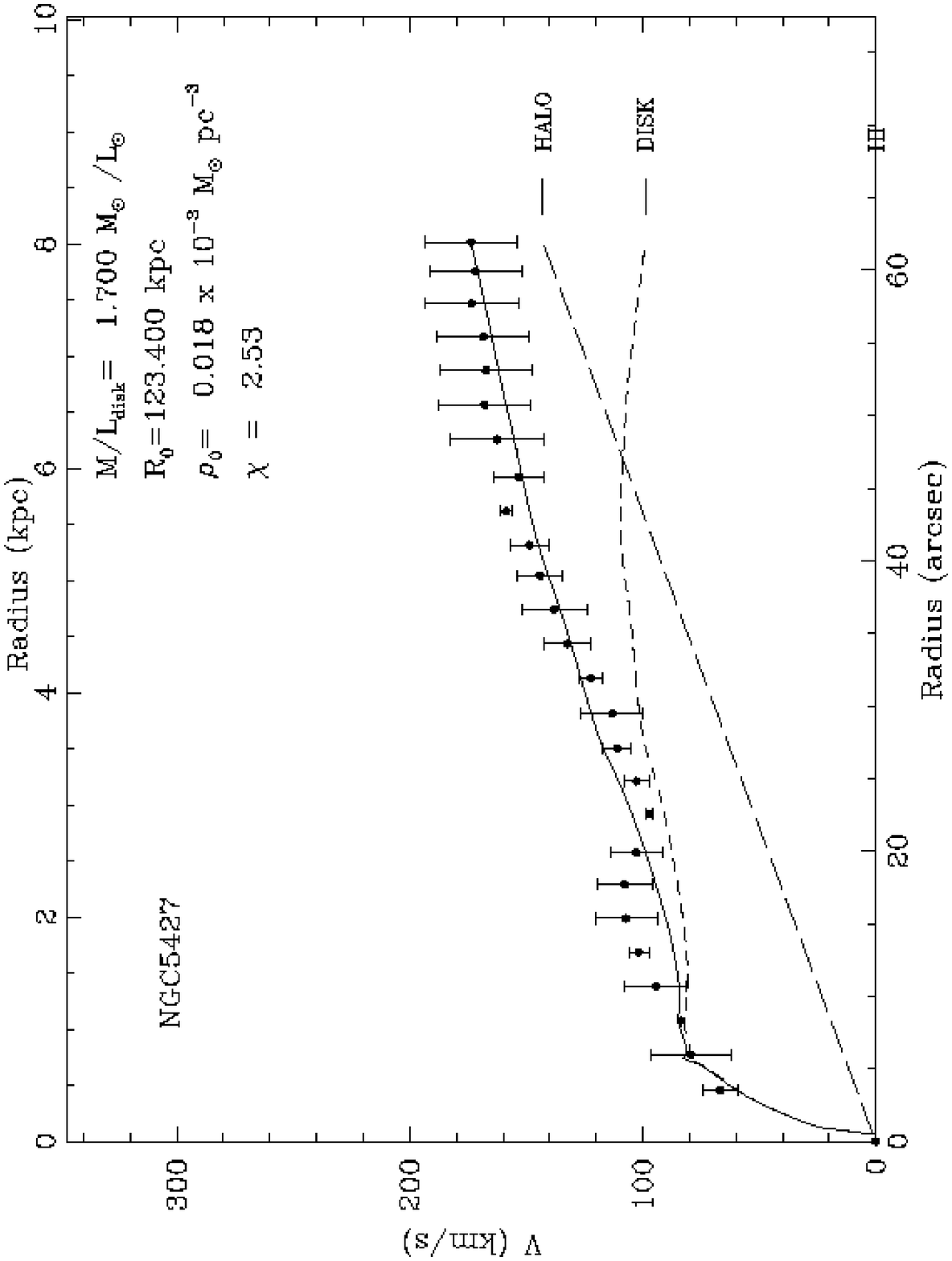}
\caption{Maximum disk mass model for NGC 5427 using an isothermal halo. The parameters displayed in each curve stand for the mass-to-light ration of the stellar disk ($M/L_{disk}$), the core radius of the dark matter halo and its central density ($R_0$ and $\rho_0$,  respectively) and the minimized $\chi^2$  in the three-dimensional parameter space. Mass-model taken from \citet{blais01}.}
\end{figure*}

The previously computed ratio $  M_{add} /  M_{orbital}$   of  $0.62$ is consistent with 
an important contribution 
of unseen matter in the outermost parts of the galaxies that cannot be acquainted through 
the computation of each galaxy's mass using the optical $H \alpha$ rotation curve and Lequeux's method. 
On the other hand the mass of NGC 5427 might have been uderestimated due to an overestimation of 
its inclination. The uncertainty in inclination determination has a big effect  on the derived velocity 
amplitude, and thus masses and $M/L$ ratios:  for example, $1 / \sin \ 34^\circ = 1.79$ and 
$1 / \sin \ 24^\circ =2.46$,  indicating already a factor of 1.9 difference in masses.
This would explain why even if within smaller radii NGC 5426 seems to be more massive, 
the radial velocities of the  bridge-like feature between both galaxies seems to belong to the latter 
implying it is more affected by the interaction than the companion NGC 5427.
The fact that probably these two galaxies have approximately the same masses  supports the idea 
of them being {\it twin} galaxies as suggested by \citet{yam89}.

\section{Discussion}

As  suggested by \citet{bla82},  the asymmetric distribution of light of the
pair  could be due to the fact that NGC 5427 has not had
enough time to average the distribution of its outer parts.
Considering that for this galaxy $V_{rot} \sim 172 \ km  \ s^{-1}$
at $R \sim 9 \ kpc$,  the rotation period would amount to $ \sim 3 \times 10^{8}  \   yrs $ 
thus sustaining \citet{bla82}'s suggestion of this being a ``recent'' encounter.
Nevertheless  \citet{scho90} showed that the bridges and tails of 
interacting galaxies are bluer than the inner disks showing that on average star formation has 
increased in the interaction zone. Also, N-body models of interacting galaxies show star formation 
is induced simultaneously in all azimuthal directions.
These two last results (found after Blackman's work) do not support the idea that the surface brightness 
in the interaction zone between the two galaxies is weaker than on the opposite sides of the galaxies, 
because the pair has not had time to average the star formation in these galaxies.
A second and more convincing explanation for the lack of emission on the adjacent halves of the galaxies was also given 
by \citet{bla82}. According to him, this lack of emission could be explained by obscuration of dust in 
NGC 5426's disk whose inclination is larger than that of NGC 5427's. This would imply NGC 5426  
is in front of NGC 5427.

NGC 5426 is a two-armed spiral galaxy with well-defined arms so we may assume it is a
trailing spiral since leading spiral arms are an uncommon phenomena \citep{pas82} and the 
dynamical theory  to explain them is mostly associated to the formation of a single leading arm 
\citep{thom89}. Considering  this sens of rotation and the position of the approaching and the 
receding radial velocities  we can infer this galaxy's orientation placing its western side
closer to the observer than the eastern side.
For NGC 5427 we can use the same argument to determine it is a trailing spiral. This spin is also 
suggested by \citet{alf01} according to the movement of the gas as it goes through the
density wave.  Considering  the distribution of approaching and receding radial velocities,
the southeastern   side of the galaxy is closer to the observer than the  northwestern one.
From both the tilt and spin of the galaxies as well as from the position of one with respect to the 
other a possible  sense of  passage of the encounter can be suggested. The  fact that NGC 5427 is 
a two-arm grand-design spiral suggests the encounter with NGC 5426 is a direct one. 
Comparing the morphology of this pair with the simulations of \citet{toom72} we find that the
bridge-like feature observed in Arp 271 resembles the elongation presented by these authors in
the -0.5 frame of their Figure 3 which corresponds to the first stages ( $ 0.5 \times 10^8 \ yrs$
before perigalactic instant) of a flat direct parabolic passage of a companion of equal mass.
Though this is a loose comparison and the system is likely to be more complicated, we think this first 
analysis can  help as  a starting point for future numerical simulations of this particular encounter.
The suggested global  geometry of the encounter as well as the sense of passage are  shown in Figure 15.

As already done by \citet{bla82}, the time scale of the interaction was estimated from the ratio 
of the projected separation between 
both galaxies ($18.17 \ kpc$) and the difference in the systemic velocities of the pair ($147.5 \ km\ s^{-1}$) 
resulting in $ \sim 10^{8}  \   yrs $, a value which is also  similar to the time-scale in the \citet{toom72} 
simulation.
This result implies  the interaction is not necessarily very strong, in agreement with the fact that the 
galaxies are not very distorted. This makes it possible to assume that the noncircular motions are not
 necessarily very strong so that it is possible thus validating the mass estimates for the galaxies.
However, there do exist characteristics that can be associated to galaxy interaction process such
as the bridge of material, the  distorted velocity field in NGC 5427 and its straight spiral arm 
segment.
The range of radial velocities observed along the straight arm   is considerably
 smaller  than in the other  parts of this galaxy's arms  as if the whole structure were slowing  down to an  average
velocity due to the proximity with NGC 5426 and/or  as if this part of the spiral structure is being deviated
from the galactic plane during the interaction.
The small  shift between the central isovelocity contours  in NGC 5426  seems to be outlining a small
 and incipient central  bar  of about $6 \arcsec $ whose formation could also have been 
triggered by the interacting process.
The giant HII regions in each galaxy could also  be sign of recent star formation triggered by the
interacting process.
Nevertheless, these peculiarities could also be due to the intrinsic nature of
spiral galaxies. For example, the origin of bars in interacting
disk galaxies remains unclear. Though there have been works
showing statistical evidence that companions trigger  the
formation of bars \citep{elm90, mar00}, bars can be formed as a
result  of disk instabilities regardless of their  environment \citep{van02}.
For example,  it was shown by \citet{hohl71} that cool  stellar disks are very sensitive to bar formation,  
the so-called ``bar instability''  scenario later decribed by 
\citet{sell93}.  On the other hand,  \citet{nog87} and \citet{miw98} 
 have shown by N-body simulations that galaxy interaction can trigger bar formation. 
Detailed numerical simulations considering strict kinematical restrictions need to be done  to establish  
the extent to  which  the interacting process  is related
to the formation of a possible nuclear bar -both in the case of NGC 5427's central bar and 
of NGC 5426's small incipient bar-like feature.

\begin{figure*}
\centering
\includegraphics{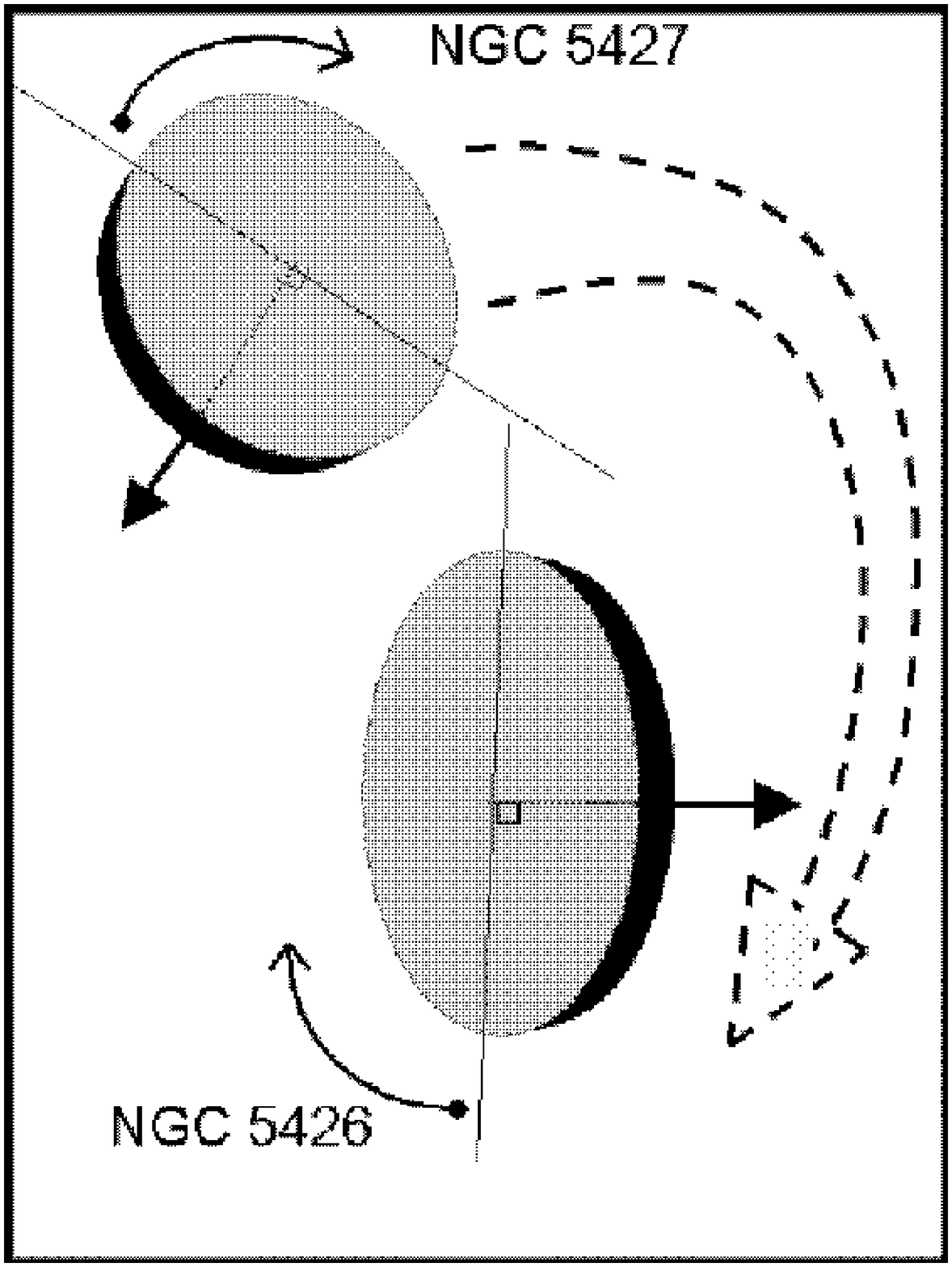}
\caption{Possible 3D  configuration for the encounter between NGC 5426 and NGC 5427. 
The spin and orientation of each galaxy is indicated. The large arrow indicates the possible sense of 
passage of one galaxy with respect to the other where NGC 5427 is moving from behind NGC 5426 towards 
the foreground in the bottom-right ocrner of the frame. }
\end{figure*}

Another interesting feature observed in Arp 271 is the fact  that the bridge  of stars 
does not appear to match or follow  any other structure in neither two galaxies. 
In fact, this bridge  is almost perpendicular to the linear
 segment in the southern arm of NGC 5427 and it does not quite seem to follow the 
western arm of NGC 5426.
 This peculiarity has been observed in other galaxies where tidal tails,  bridges and spiral structure
 coexist, yet do not seem to join smoothly. In systems such as Arp 96 and  M100, the disk's spiral
 structure appears to have begun to decouple from the tidal structure \citep{schw98}.
These differences between gas and stars could be due to
non-gravitational forces such as  ram-pressure due to hot gas in
quiescent halos, starburst-driven outflows or magnetic fields.
In our case, this could also be some part of the spiral structure deviated from the
galactic plane during the interaction.
 The study  of   galaxies showing this peculiarity can give us information  about the kinematical and
dynamical processes  that lead to the formation  and sustainment of grand design spiral structure and the
parameters that control it.
The fact that the systemic velocities differ by almost $150 \  km \  s^{-1} $ and   that the galaxies 
seem to be connected by a bridge of material suggests that we are actually measuring an important 
component of the peculiar velocities between these two galaxies.
This relative radial velocity difference might represent a significant component of the
total velocity difference. So,  if we assume  for a moment that the
sky-plane velocity components are exactly zero, NGC 5426
would ``see'' the  other galaxy passing away from observer, thus in
an almost retrograde sense with respect to its internal rotation 
(orbital inclination about 150 degrees), and similarly NGC 5427
would see an almost perpendicular passage of the other galaxy.
The almost perpendicular passage seen by NGC 5427 would be consistent
with the absence of apparent tails or bridges emanating from it
-but might still help to excite spiral structure.
The almost retrograde passage seen by NGC 5426 might explain the weakness of bridges.

\section{Conclusions}

The detailed study of the kinematics and dynamics of interacting galaxy pairs provides important information
about  the structure of each member of the pair as well as of the  encounter as a whole.
In this work we presented Fabry-Perot observations of the isolated interacting galaxy pair NGC 5426/27 (Arp 271) showing that for a perturbed system it is important to have kinematical information of large portions of the galaxies studied.
From the analysis of the velocity field and the computation of the FWHM of the intensity profiles and residual velocities maps, we determined  non-circular motions in each galaxy and found them to be possibly  associated with the passage of gas through the spiral arms and also with the flow of gas into or out of the central parts of each galaxy along bar-like features.  
From the observed rotation curve of each galaxy and the fit of different mass models, 
we found that though NGC 5426 seems to be  more massive than NGC 5427 within its optical radius, both galaxies seem to have
approximately the same total mass, thus belonging to the {\it twin} galaxies scheme.
For NGC 5427 the observed rotation curve can be fitted with a maximum disk ($M/L=1.7$) and a pseudo-isothermal halo.
Though this seems to be a relatively early encounter,  several morphological features of each galaxy were associated with the interaction process, such as an incipient bar-like feature in the central parts of NGC 5426 and a straight segment in one of NGC 5427's arms. The radial velocities of the HII regions outlining a bridge between the two galaxies were also derived finding that this structure seems to be associated with NGC 5426.
 Finally from the resulting kinematics and dynamics, we suggest a possible 3D configuration for this encounter. This will be used as a starting point in future  numerical simulations  of an encounter of this sort.

\begin{acknowledgements}
      We thank the staff of the OAN (Observatorio Astron\'omico Nacional) for their support during PUMA data acquisition. We acknowledge the financial support from grant IN 104696 of DGAPA-UNAM. We also acknowledge C. Carignan for allowing us to use his mass model. I. Fuentes-Carrera acknowledges financial support from Fundaci\'on UNAM and  CONACYT and DGEP grants. I. Fuentes.Carrera also thanks the Observatoire de Marseille for its hospitality during several research stays with financial support from ECOS(France)/ANUIES(Mexico).
\end{acknowledgements}

\bibliographystyle{aa}

\end{document}